\documentclass[pdflatex,aps,prd,twocolumn,nofootinbib,showpacs,superscriptaddress]{revtex4-1}
\usepackage{graphicx} \usepackage{subfigure} \usepackage{lipsum}
\newcommand{\nc}[1]{\newcommand{#1}} \newcommand{\be}{\begin{eqnarray}}
	\newcommand{\ee}{\end{eqnarray}} 
\nc{\hmu}{\hat{\mu}}
\nc{\bmu}{\bar{\mu}}
\nc{\bchi}{\bar{\chi}}
\def\lsim{\raise0.3ex\hbox{$<$\kern-0.75em\raise-1.1ex\hbox{$\sim$}}}
\def\gsim{\raise0.3ex\hbox{$>$\kern-0.75em\raise-1.1ex\hbox{$\sim$}}}
\usepackage{color}
\usepackage{epsfig}
\usepackage{graphics}
\usepackage{amssymb}
\usepackage{amsmath}
\usepackage{hyperref}
\nc{\fb}{\color{blue}}
\nc{\fr}{\color{red}}

\begin{document}

\title{Skewness, kurtosis and the fifth and sixth order cumulants of net 
baryon-number distributions from lattice QCD confront high-statistics 
STAR data
}
\author{
A. Bazavov}
\affiliation{Department of Computational Mathematics, Science and Engineering 
and Department of Physics and Astronomy, Michigan State University, East 
Lansing, MI 48824, USA}
\author{
D. Bollweg}
\affiliation{
Fakult\"at f\"ur Physik, Universit\"at Bielefeld, D-33615 Bielefeld,
Germany}
\author{
H.-T. Ding}
\affiliation{Key Laboratory of Quark \& Lepton Physics (MOE) and Institute
of Particle Physics, Central China Normal University, Wuhan 430079, China}
\author{
P. Enns}
\affiliation{
Fakult\"at f\"ur Physik, Universit\"at Bielefeld, D-33615 Bielefeld,
Germany}
\author{
J. Goswami}
\affiliation{
Fakult\"at f\"ur Physik, Universit\"at Bielefeld, D-33615 Bielefeld,
Germany}
\author{
P.~Hegde} 
\affiliation{Center for High Energy Physics, Indian Institute of Science, 
Bangalore 560012, India}
\author{
O. Kaczmarek}
\affiliation{Key Laboratory of Quark \& Lepton Physics (MOE) and Institute
of Particle Physics, Central China Normal University, Wuhan 430079, China}
\affiliation{
Fakult\"at f\"ur Physik, Universit\"at Bielefeld, D-33615 Bielefeld,
Germany}
\author{
F. Karsch}
\affiliation{
Fakult\"at f\"ur Physik, Universit\"at Bielefeld, D-33615 Bielefeld,
Germany}
\author{
R. Larsen}
\affiliation{
Physics Department, Brookhaven National Laboratory, Upton, NY 11973, USA}
\author{
Swagato Mukherjee} 
\affiliation{
Physics Department, Brookhaven National Laboratory, Upton, NY 11973, USA}
\author{\\ H. Ohno}
\affiliation{Center for Computational Sciences, University of Tsukuba,
Tsukuba, Ibaraki 305-8577, Japan}
\author{
P. Petreczky}
\affiliation{
Physics Department, Brookhaven National Laboratory, Upton, NY 11973, USA}
\author{
C. Schmidt} 
\affiliation{
Fakult\"at f\"ur Physik, Universit\"at Bielefeld, D-33615 Bielefeld,
Germany}
\author{S. Sharma}
\affiliation{
Department of Theoretical Physics, The Institute of Mathematical Sciences, 
Chennai 600113, India\\[2mm]
{\bf (HotQCD Collaboration)}}
\author{
P. Steinbrecher}
\affiliation{
Physics Department, Brookhaven National Laboratory, Upton, NY 11973, USA}
\date{\today}

\begin{abstract}

We present new results on up to $6^{th}$ order 
cumulants of net baryon-number fluctuations at small values
of the baryon chemical potential, $\mu_B$, obtained in lattice QCD 
calculations with physical values of light and strange quark masses.
Representation of the Taylor expansions of higher order cumulants 
in terms of the ratio of the two lowest order cumulants, 
$M_B/\sigma_B^2=\chi_1^B(T,\mu_B)/\chi_2^B(T,\mu_B)$,
allows for a parameter free comparison with data on net proton-number
cumulants obtained by the STAR Collaboration in the Beam Energy Scan
at RHIC. We show that recent high statistics data on skewness and 
kurtosis ratios of net proton-number distributions, obtained at
beam energy $\sqrt{s_{_{NN}}}=54.4$~GeV, agree well with lattice QCD 
results on cumulants of net baryon-number fluctuations close to the 
pseudo-critical temperature, $T_{pc}(\mu_B)$, for the chiral transition in QCD. 
We also present first results from 
a next-to-leading order expansion of $5^{th}$ and $6^{th}$ order 
cumulants on the line of pseudo-critical temperatures.
\end{abstract}

\maketitle

\section{Introduction}

The phase diagram of strong interaction matter at non-zero
temperature and non-zero baryon-number density is being 
explored intensively through numerical calculations performed in
the framework of lattice regularized Quantum Chromo Dynamics (QCD)
\cite{Ding:2015ona}, as 
well as through ultra-relativistic heavy ion collisions with varying
beam energies \cite{Luo:2017faz}. 
At vanishing and small values of the chemical potentials for conserved 
charges (baryon number ($\mu_B$), electric charge ($\mu_Q$), strangeness 
($\mu_S$))
it is well established that the transition from the low temperature hadronic 
region to the quark-gluon plasma at high temperature is a smooth transition \cite{Aoki:2006we}
characterized by a pseudo-critical temperature, $T_{pc}(\mu_B)$
\cite{Bonati:2015bha,Cea:2015cya,Bellwied:2015rza,Bazavov:2018mes}.
At larger values of the baryon chemical potential it, however, is generally expected 
that a line of first order phase transition exists, which ends in a second 
order critical point \cite{CEP,CP}.
This elusive critical point is searched for in the Beam Energy Scan (BES)
performed at the Relativistic Heavy Ion collider (RHIC) at
Brookhaven National Laboratory (BNL) \cite{Bzdak:2019pkr}. 
However, its existence as a fundamental
property of the theory of strong interactions (QCD)
still awaits confirmation.

The pseudo-critical line, $T_{pc}(\mu_B)$, which distinguishes the 
low and high temperature regimes of strong interaction matter as described
by QCD, 
has been determined quite accurately in lattice QCD calculations 
for baryon chemical potentials up to about twice the
pseudo-critical temperature, $\mu_B\ \lsim\ 2 T_{pc}(0)\simeq 300$~MeV 
\cite{Bonati:2015bha,Cea:2015cya,Bellwied:2015rza,Bazavov:2018mes}. In our 
recent analysis we found \cite{Bazavov:2018mes}
\begin{equation}
	T_{pc}(\mu_B)=T_{pc}^0 \left( 1- \kappa_2^B \left(\frac{\mu_B}{T}\right)^2 
	+ {\cal O} (\mu_B^4) \right)\; ,
\label{Tpc}
\end{equation}
with $T_{pc}^0=(156.5\pm 1.5)$~MeV and $\kappa_2^B=0.012(4)$ with a 
${\cal O} (\mu_B^4))$ correction that vanishes within errors.
At $\mu_B=0$ the pseudo-critical temperature
turns out to be in good agreement with the freeze-out temperature
determined by the ALICE Collaboration at the LHC \cite{Andronic:2017pug} 
and the pseudo-critical line, $T_{pc}(\mu_B)$, also is consistent 
with freeze-out temperatures determined by the STAR Collaboration during 
the first BES at RHIC (BES-I) \cite{Adamczyk:2017iwn},
albeit these temperatures have larger statistical errors. 

The experimental determination of freeze-out parameter is based on
a measurement of particles yields, {\it i.e.} first moments of particle 
distributions, which in turn are closely related to first order cumulants
of net charge fluctuations. The proximity of freeze-out temperatures and
the pseudo-critical temperature determined in QCD suggests that also
the higher order moments of net charge fluctuations reflect properties
of a thermal medium close to the pseudo-critical line. 
This, however, is not at all well established and many caveats have been
discussed suggesting that the relation of higher order
cumulants, measured experimentally, to cumulants of conserved
charge fluctuations, calculated in equilibrium QCD thermodynamics,
is not at all straightforward \cite{Bzdak:2019pkr,Ratti}. 

Higher order cumulants of net conserved charge fluctuations are obtained as
derivatives of the logarithm of the QCD partition functions with
respect to the chemical potentials of conserved charges, 
$\vec{\mu} =(\mu_B,\mu_Q,\mu_S)$,
\begin{equation}
	\chi_n^X(T,\vec{\mu}) = \frac{1}{VT^3}
	\frac{\partial^n \ln Z(T,\vec{\mu})}{\partial \hat{\mu}_X^n}\;\; ,
	\;\; X=B,\ Q,\ S\; ,
\label{cumulants}
\end{equation}
with $\hat{\mu}\equiv \mu/T$.
These higher order derivatives become increasingly
sensitive to long range correlations and large fluctuations
in the vicinity of a critical point. At least from the theoretical point of 
view higher order cumulants thus are ideally suited to search for a 
possible critical point in the QCD phase diagram 
\cite{Hatta:2003wn,Stephanov:2008qz,Friman:2011pf}. 
The BES at RHIC
aims at finding evidence for such a critical point through the analysis
of e.g. higher order cumulants of net proton-number fluctuations which are
considered to be good proxies for cumulants of net baryon-number 
fluctuations. Results, obtained with BES-I at RHIC, indicate that qualitative 
changes in the behavior of net proton-number fluctuations occur at beam 
energies $\sqrt{s_{_{NN}}}\sim 20$~GeV \cite{STARp08,Adam:2020unf}. 
This may hint at the existence of a critical point for large values of the 
baryon chemical potential. 

While the finding of non-monotonic behavior of higher order cumulants of
net proton-number fluctuations generated a well justified
excitement \cite{STARp08,Adam:2020unf}, we still need to establish that this 
behavior is caused by
thermal fluctuations in the vicinity of a critical point and that these
higher order cumulants indeed probe thermal conditions at the time of freeze-out. 
As pointed out in Ref.~\cite{Bazavov:2017tot} 
at least for small values of the baryon chemical potential 
the first four cumulants 
of net baryon-number fluctuations, {\it i.e.} mean 
($M_B\equiv \chi_1^B(T,\vec{\mu})$), 
variance ($\sigma_B^2=\chi_2^B(T,\vec{\mu})$),
skewness ($S_B=\chi_3^B(T,\vec{\mu})/\chi_2^B(T,\vec{\mu})^{3/2}$) 
and kurtosis ($\kappa_B=\chi_4^B(T,\vec{\mu})/\chi_2^B(T,\vec{\mu})^2$) 
are predicted in QCD equilibrium thermodynamics to be related.
For $\mu_S=\mu_Q=0$ one finds
\begin{eqnarray}
	\kappa_B \sigma_B^2~~ &<& S_B\sigma_B^3/M_B \; , \nonumber \\
\Leftrightarrow\;\;	
\frac{\chi_4^B(T,\vec{\mu})}{\chi_2^B(T,\vec{\mu})}  &<&
	~\frac{\chi_3^B(T,\vec{\mu})}{\chi_1^B(T,\vec{\mu})} \; .
	\label{ordering}
\end{eqnarray}
This relation, which is only slightly violated in strangeness neutral 
systems, has been established in lattice QCD calculations
using next-to-leading order (NLO) Taylor expansions of the first four cumulants
of net baryon-number fluctuations \cite{Bazavov:2017tot}.
The data on cumulants of net-proton number fluctuations, obtained by STAR 
during BES-I \cite{Adam:2020unf} at beam energies 
$\sqrt{s_{_{NN}}}\ge 19.6$~GeV are, on average, consistent with this 
finding \cite{Bazavov:2017tot}.
However, statistical errors are large and, for instance, data obtained
at $\sqrt{s_{_{NN}}}=62.4$~GeV violate the above relation. Results at
several other beam energies are inconclusive due to the large statistical
errors on the fourth order cumulant ratio $\kappa_B\sigma_B^2$. 
To this extent recent high statistics data obtained by the STAR 
Collaboration at $\sqrt{s_{_{NN}}}=54.4$~GeV \cite{STAR54} are encouraging. 
As will be discussed in Section IV, these data fulfill the above inequality 
and the difference of the cumulant ratios 
given in Eq.~\ref{ordering} agrees 
with lattice QCD results even on a quantitative level.

We will present here new results on the density dependence of
up to $6^{th}$ order cumulants of net baryon-number fluctuations.
We calculate Taylor series at non-zero values of the baryon-number, 
electric-charge and strangeness chemical potentials that involve up
to $8^{th}$ order cumulants. We perform
these expansions for the case of strangeness neutral systems, $n_S=0$,
with a ratio of electric-charge to baryon-number, $n_Q/n_B=0.4$, that is 
representative for the conditions met in heavy ion collisions.
This allows to construct Taylor expansions
for $n^{th}$ order cumulants\footnote{Rather than specifying in the
argument of $\chi_n^B$ all three chemical potentials, $\vec{\mu}$,
we give in the strangeness neutral case only the baryon chemical potential.}, 
$\chi_n^B(T,\mu_B)$, up to ${\cal O}(\mu_B^{8-n})$. 

For the case of the skewness and kurtosis ratios,
$S_B\sigma_B^3/M_B$ and $\kappa_B\sigma_B^2$, respectively,
we thus can extend earlier NLO calculations and perform 
next-to-next-to-leading order (NNLO) expansions
that allow to better control truncation effects in the Taylor
series. We also present, for the first time, results from NLO calculations 
for the hyper-skewness and hyper-kurtosis ($5^{th}$ and  $6^{th}$ order cumulants)
ratios $\chi_5^B(T,\mu_B)/\chi_1^B(T,\mu_B)$ and
$\chi_6^B(T,\mu_B)/\chi_2^B(T,\mu_B)$. We show that these ratios
are expected to be negative at $\sqrt{s_{_{NN}}}=54.4$~GeV, in contrast 
to the preliminary findings for $6^{th}$ order cumulants of net proton-number
fluctuations reported by the STAR Collaboration \cite{STAR54}.

This paper is organized as follows. In the next section we briefly
present our calculational setup, the new statistics collected on
lattices of size $32^3\times 8$ and $48^3\times 12$ and the general
fitting ansatz used for fits at fixed values of $N_\tau = 8$ and $12$,
joint fits of these data as well as continuum limit estimates.
In section III we present results for Taylor expansions of cumulants of
net baryon-number fluctuations that use up to $8^{th}$ order cumulants.
We compare these results with experimental data for cumulants of net
proton-number fluctuations in Section IV. Section V contains our 
conclusions. Explicit expressions for the first four Taylor expansion
coefficients of net baryon-number cumulants are given in an appendix.

\section{Calculational setup}

\begin{table}[tb]
\begin{center}
\vspace{0.3cm}
\begin{tabular}{|c|r||c|c|}
\hline
\multicolumn{2}{|c||}{$N_\tau=8$}&\multicolumn{2}{|c|}{$N_\tau=12$} \\
\hline
 T[MeV] & \#conf.~~~&T[MeV] & \#conf. \\
\hline
134.64&1,275,380&134.94&256,392 \\
140.45&1,598,555&140.44&368,491 \\
144.95&1,559,003&144.97&344,010 \\
151.00&1,286,603&151.10&308,680 \\
156.78&1,602,684&157.13&299,029 \\
162.25&1,437,436&161.94&214,671 \\
165.98&1,186,523&165.91&156,111 \\
171.02&373,644&170.77&144,633 \\
175.64&294,311&175.77&131,248 \\
\hline
\end{tabular}
\end{center}
\caption{Number of gauge field configurations on lattices of size
$32^3\times 8$ and $48^3\times 12$ used
in the analysis of up to $8^{th}$ order Taylor expansion coefficients.
The values of the gauge coupling as well as the strange and light quark mass
parameter at these temperature values are taken from \cite{Bazavov:2017dus},
where also details on the statistics available on the $24^3\times 6$
lattices are given. All configurations are separated by 10 time units
in RHMC simulations \cite{Bazavov:2017dus}.
}
\label{tab:statistics}
\end{table}

Up to $4^{th}$ order cumulants of net baryon-number fluctuations have been 
calculated previously \cite{Bazavov:2012vg,Borsanyi:2014ewa,Bazavov:2017tot} 
in a next-to-leading order Taylor expansion. In particular, we performed 
calculations \cite{Bazavov:2017tot} with the Highly Improved 
Staggered Quark (HISQ) \cite{Follana:2006rc} discretization 
scheme for $(2+1)$-flavor QCD with a physical strange quark mass and two
degenerate, physical light quark masses.  
Here we extend these calculations by increasing the number of gauge field
configurations generated on lattices of size $32^3\times 8$ and
$48^3\times 12$ by a factor $3$-$5$ in the transition region and at least
a factor 2 at other values of the temperature. This allows us to
calculate up to $8^{th}$ order cumulants of net baryon-number, net strangeness
and net electric-charge fluctuations, including  also their correlations,
at vanishing values of the chemical
potentials. These cumulants provide expansion coefficients in Taylor series
for net baryon-number cumulants $\chi_n^B(T,\vec{\mu})$. We 
calculate NLO expansions for $5^{th}$ and $6^{th}$ order 
cumulants and obtain NNLO results for $3^{rd}$ and $4^{th}$ order
cumulants. In the case of $1^{st}$ and $2^{nd}$ order cumulants,
{\it i.e.} the mean and variance of net baryon-number distributions,
we even obtain NNNLO results. The set of gauge field ensembles, 
which has been used in this analysis, and the
number of gauge field configurations per ensemble
on lattices with temporal extent $N_\tau=8$ and $12$ are summarized 
in Table~\ref{tab:statistics}. 

\begin{figure}[t]
\begin{center}
\includegraphics[width=72mm]{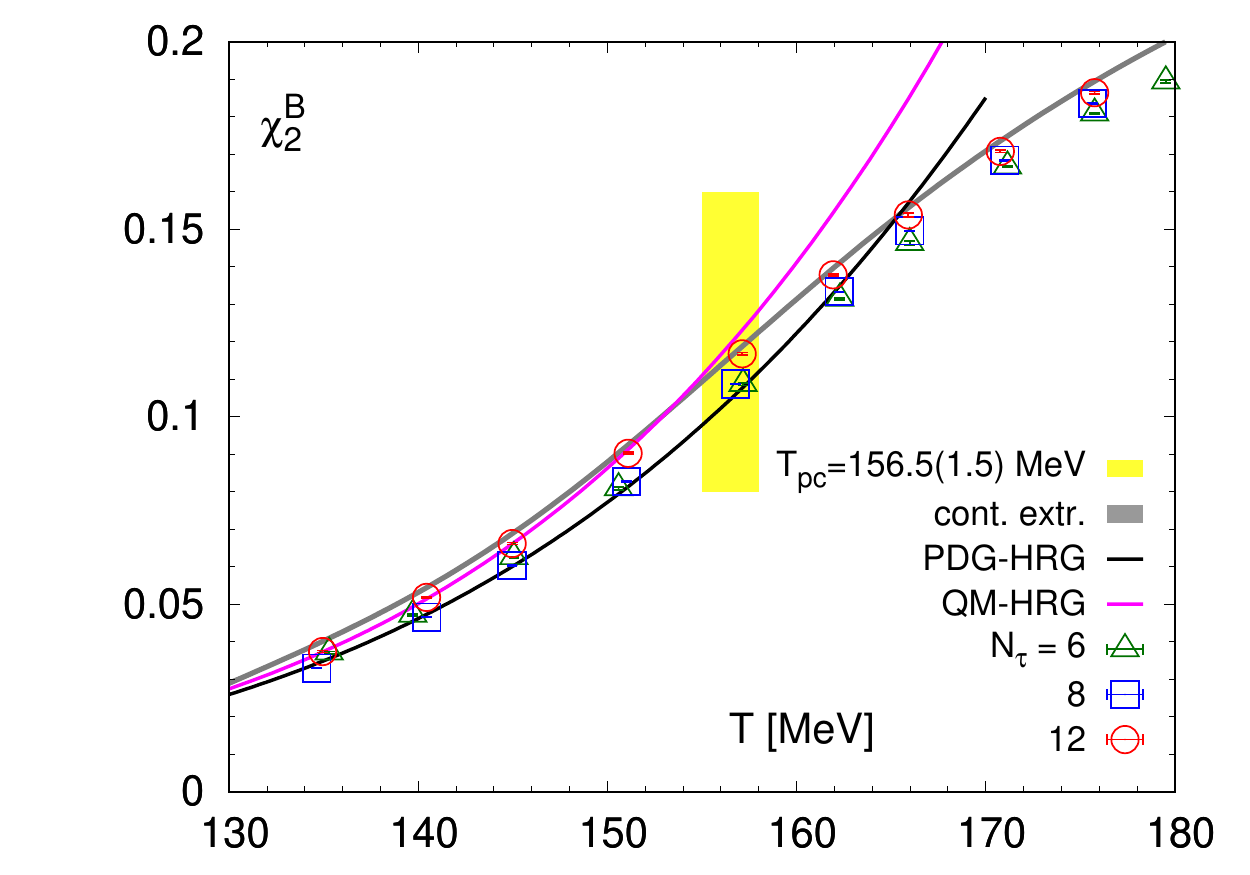}
\vspace*{-0.3cm}
\includegraphics[width=72mm]{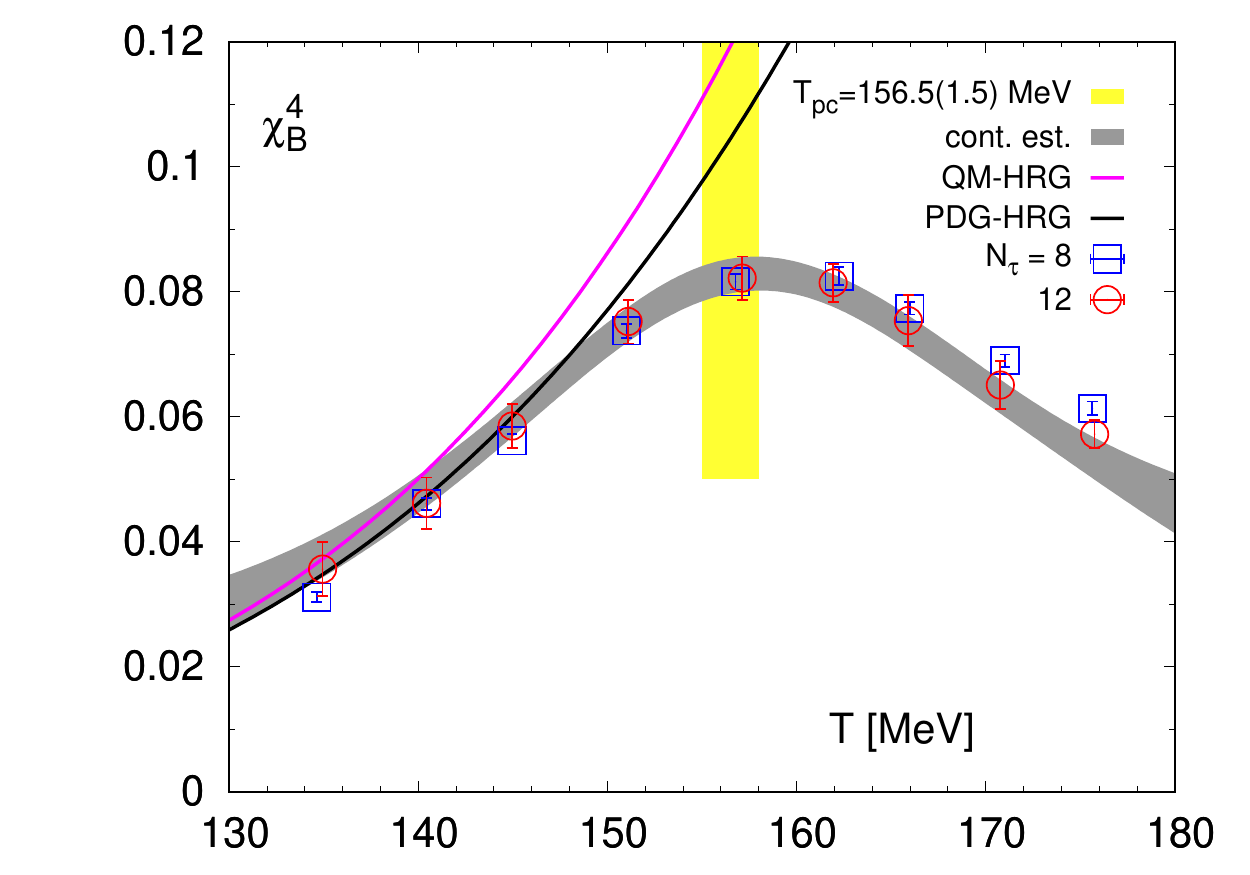}
\vspace*{-0.3cm}
\includegraphics[width=72mm]{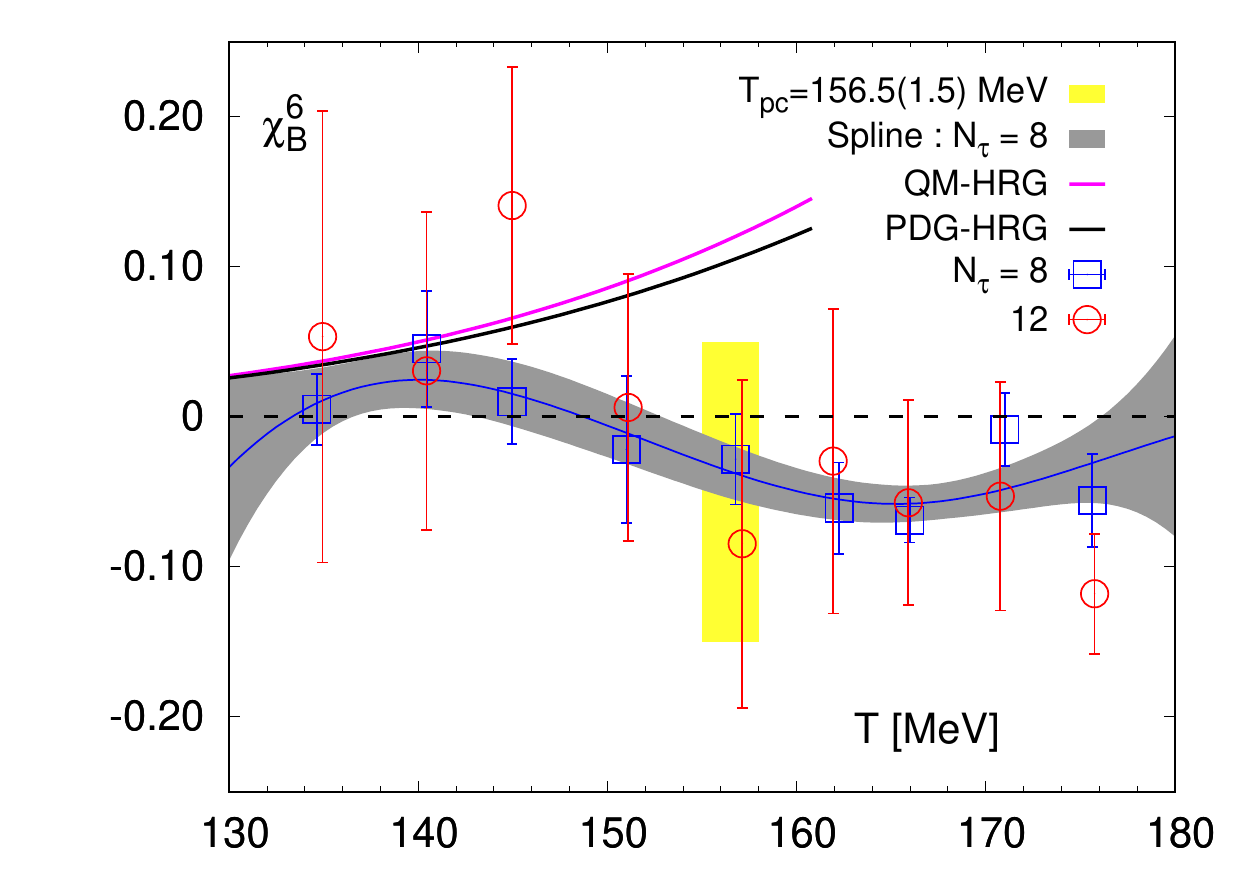}
\vspace*{-0.2cm}
\hspace*{1mm}\includegraphics[width=72mm]{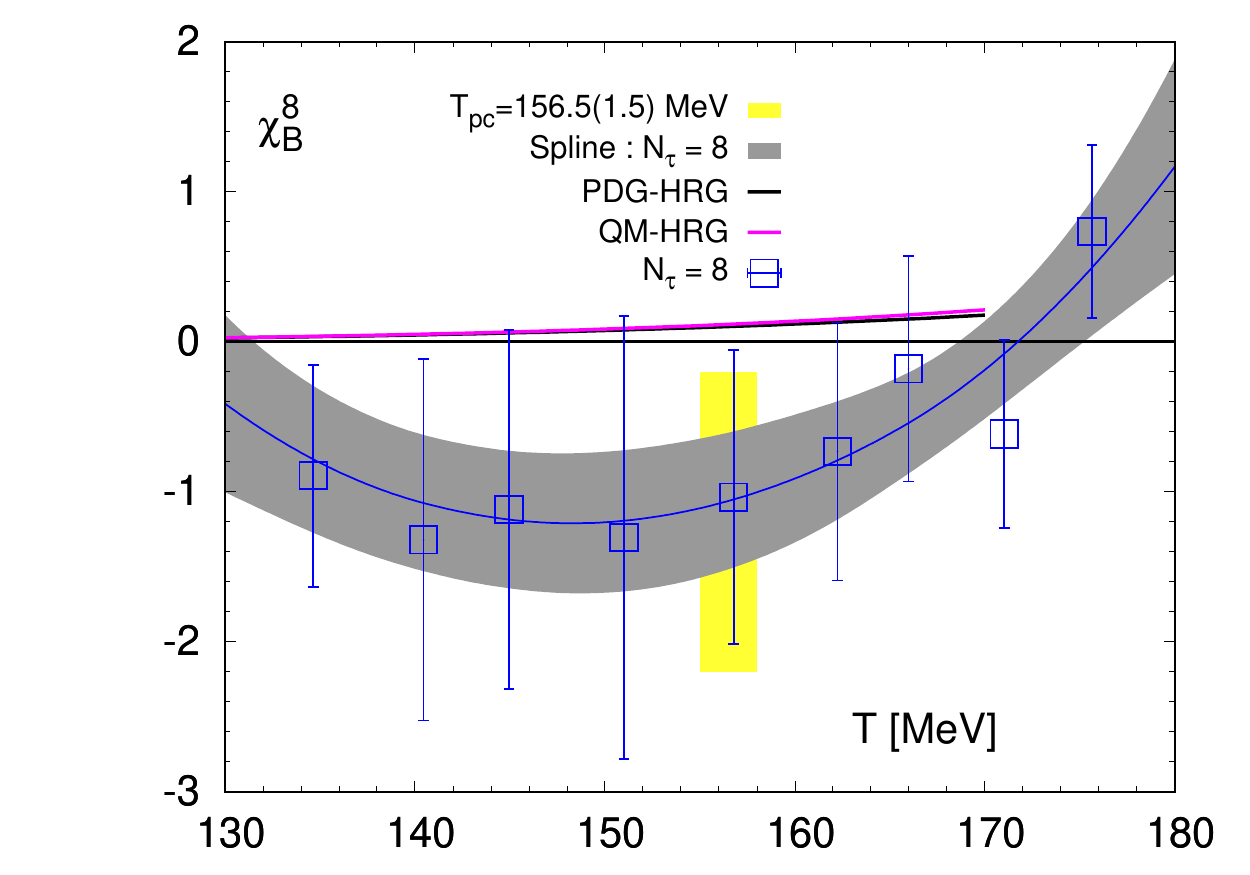}
\vspace*{-0.2cm}
\caption{Cumulants of net baryon-number fluctuations
        from second to eighth order (top to bottom)
evaluated at $\mu_B=0$ on lattices of size $N_\sigma^3\times N_\tau$
with $N_\sigma = 4N_\tau$. For further details see text.
}
\vspace*{-1.0cm}
\label{fig:Bdiag}
\end{center}
\end{figure}

Results for up to $8^{th}$ order diagonal 
net baryon-number susceptibilities, $\chi_n^B\equiv \chi_n^B(T,0)$, 
are given in Fig.~\ref{fig:Bdiag}.
For the quadratic fluctuations, $\chi_2^B$, we also show results for 
lattices with temporal extent $N_\tau=6$, which already
had been used in \cite{Bazavov:2018mes}. For the $8^{th}$ order cumulant,
$\chi_8^B$, we only show our results for $N_\tau=8$ as statistical errors
on the $N_\tau=12$ data are still too large. 
The bands shown in these
figures give a continuum extrapolation for  $\chi_2^B(T)$ using data
from calculations for three different lattice spacings ($aT=1/N_\tau$) and
a continuum estimate for $\chi_4^B(T)$ based on $N_\tau=8$ and $12$ data sets.
For $\chi_6^B(T)$ and $\chi_8^B(T)$ we only show spline interpolations
of the data obtained on the $32^3\times 8$ lattices. 
Results for these cumulants, obtained from calculations within a
non-interacting hadron resonance gas (HRG) model that use resonances from
the particle data tables \cite{PDG} (PDG-HRG) as well as additional resonances 
calculated within the Quark Model \cite{Isgur,Ebert} (QM-HRG) are given by 
lines. The latter list contains
additional resonances not (yet) observed experimentally.

We determine the expansion coefficients, $\tilde{\chi}_n^{B,k}(T)$, 
for Taylor series of $n^{th}$ order cumulants,
\begin{equation}
	\chi_n^B(T,\mu_B) = \sum_{k=0}^{k_{max}} \tilde{\chi}_n^{B,k}(T) 
	\hat{\mu}_B^k
	\; ,
	\label{chin}
\end{equation}
for the case of vanishing net strangeness 
density, $n_S=0$, and an electric-charge to baryon-number ratio, 
$n_Q/n_B=0.4$. Explicit expressions for the NLO expansion coefficients 
of up to $6^{th}$ order net baryon-number cumulants are given in 
\cite{Bazavov:2017tot}.
The explicit form of the higher order expansion coefficients are
given in the appendix.

Using the Taylor series for $n^{th}$ order cumulants, Eq.~\ref{chin}, we 
construct cumulant ratios with polynomials of order $[k_{max},l_{max}]$,
\begin{eqnarray}
	R_{nm}^B  &=& \frac{\chi_n^B(T,\mu_B)}{\chi_m^B(T,\mu_B)} 
= \frac{\sum_{k=1}^{k_{max}} \tilde{\chi}_n^{B,k}(T) \hat{\mu}_B^k}{\sum_{l=1}^{l_{max}} \tilde{\chi}^{B,l}_m(T) \hat{\mu}_B^l} \; . 
\label{ratios}
\end{eqnarray}
In order to control systematic effects arising from the truncation of the 
Taylor series expansion for the cumulant ratios $R_{nm}^B$, we calculate these
ratios using different orders of the Taylor expansion for the cumulants
appearing in the numerator and denominator of these ratios. We analyzed 
the polynomial ratios for different $[k_{max},l_{max}]$ as well as Taylor 
expansions of the ratios themselves. We find that the former are more stable 
at large $\mu_B/T$. In the following we will use the ratios of polynomials 
with $[k_{max},l_{max}]$ corresponding to identical orders 
(LO, NLO, NNLO, NNNLO) of expansions in the cumulants appearing in the 
numerator and denominator, respectively.

We fit cumulant ratios using a rational polynomial ansatz,
\begin{equation}
	f(T,\hat{\mu}_B) = \frac{\sum_{n=0}^{n_{max}} a_{n}(\hat{\mu}_B)
	\bar{T}^n}{\sum_{m=0}^{m_{max}} b_{m} (\hat{\mu}_B)
	\bar{T}^m} \;\; ,\;\; {\rm with} \;\; \bar{T}=\frac{T}{T_0}\; ,
        \label{fitansatz}
\end{equation}
where $T_0$ is some arbitrary scale.
When using this rational polynomial ansatz for fits at non-zero
$\mu_B$ we allow for a quadratic $\mu_B$-dependence of all 
expansion coefficients, $a_{n}(\hat{\mu}_B) = a_{n,0}+a_{n,2}\hat{\mu}_B^2$ and 
similarly for $b_n(\hat{\mu}_B)$. When performing joint fits of data 
on lattices with different sizes and lattice spacings, $a$, we allow for 
${\cal O}(a^2)$ cut-off corrections 
that are parametrized in terms of the temporal lattice extent $N_\tau=1/aT$, 
e.g.
\begin{equation}
f(T,\hat{\mu}_B) = h(T,\hat{\mu}_B) + \frac{1}{N_\tau^2} g(T,\hat{\mu}_B) \; ,
\label{joint-fit}
\end{equation}
with $g(T,\hat{\mu}_B)$ and $h(T,\hat{\mu}_B)$ being rational 
polynomials of the type given in Eq.~\ref{fitansatz}.

\section{Cumulants of net baryon-number fluctuations}

\subsection{Mean and variance of net baryon-number fluctuations}
\begin{figure}[t]
\begin{center}
\includegraphics[width=85mm]{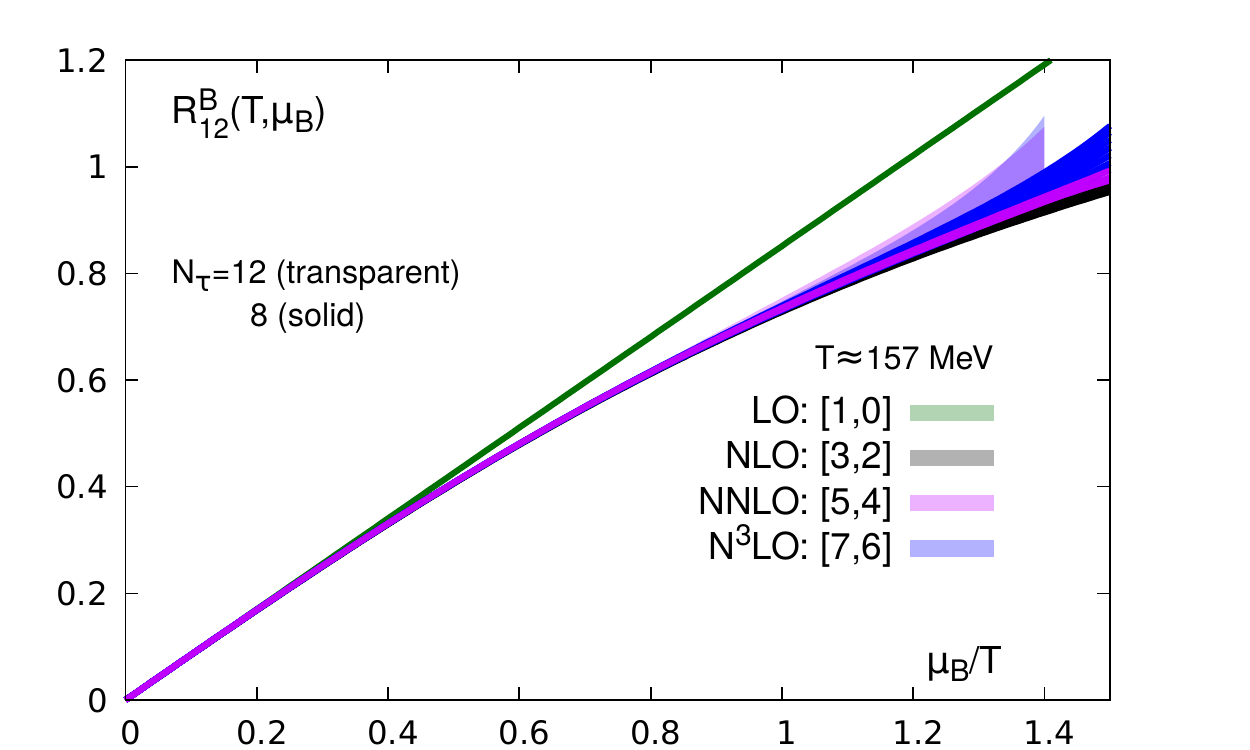}
\caption{Expansion of
        $R_{12}^B\equiv M_B/\sigma_B^2$ at a fixed temperature close to
        the pseudo-critical line
        $T_{pc}(\mu_B)$ versus the baryon chemical potential.
        Shown are results from up to N$^3$LO expansions on lattices
        of size $32^3\times 8$ and $48^3\times 12$.
}
\label{fig:R12B}
\end{center}
\end{figure}

We have calculated the ratio of mean, $M_B=\chi_1^B(T,\mu_B)$, and variance, 
$\sigma_B^2=\chi_2^B(T,\mu_B)$, of net baryon-number fluctuations, 
\begin{equation}
	R_{12}^B(T,\mu_B)\equiv \frac{M_B}{\sigma_B^2} = 
	\frac{\chi_1^B(T,\mu_B)}{\chi_2^B(T,\mu_B)}\; ,
\label{R12B}
\end{equation}
for systems with vanishing net strangeness, $n_S=0$, and a net electric-charge 
to net baryon-number density $n_Q/n_B=0.4$ on lattices with
temporal extent  $N_\tau =8$ and $12$. Using up to $8^{th}$ order Taylor
expansion coefficients, we can construct Taylor series
up to order ${\cal O}( \hat{\mu}_B^7)$ and 
${\cal O}(\hat{\mu}_B^6)$ for $\chi_1^B(T,\mu_B)$ and
$\chi_2^B(T,\mu_B)$, respectively. Truncating these series at $k_{max}$
and $l_{max}=k_{max}-1$, respectively, we construct the $[k_{max},l_{max}]$
polynomial ratios which provide leading order ($[1,0]$, LO), next-to-leading 
order  ($[3,2]$, NLO) etc approximations for the ratio of mean  
and variance of the distribution for net baryon-number 
fluctuations, $R_{12}^B\equiv M_B/\sigma_B^2$. Results for different 
$[k_{max},l_{max}]$ are shown in Fig.~\ref{fig:R12B}. The figure shows 
results obtained  
on lattices with temporal extent $N_\tau=8$ and $12$ at 
a temperature\footnote{As is evident from Tab.~\ref{tab:statistics} the 
temperatures differ slightly for the two lattice sizes, 
$T=156.76$~MeV for $N_\tau=8$ and $T=157.13$~MeV for $N_\tau=12$, 
respectively.} $T\simeq 157$~MeV
which is close to the pseudo-critical temperature at $\mu_B=0$.

We find that cut-off effects are negligible for $\mu_B/T\le 1$ and
remain comparable to the statistical errors for the 
$N_\tau=12$ data at least up to $\mu_B/T\simeq 1.2$. This holds true in the
entire temperature range $T\in [135~{\rm MeV}:175~{\rm MeV}]$ analyzed by us. 
Differences in $R_{12}^B$ constructed from NNLO and NNNLO Taylor series
of the cumulants are  about 2\% for $\mu_B/T=1$.

As the temperature dependence of $R_{12}^B$ is weak in the temperature
range considered by us and also deviations of the $\mu_B$-dependence from
the leading order, linear behavior are moderate we find that using $[2,3]$ 
rational polynomials in both terms of the fit ansatz given
in Eq.~\ref{joint-fit} are sufficient for obtaining good fits to the data. 
We performed fits separately for the NNLO and NNNLO data sets at fixed
values of $T$ and $\mu_B/T\le 1.2$.
\begin{figure}[t]
\begin{center}
\includegraphics[width=85mm]{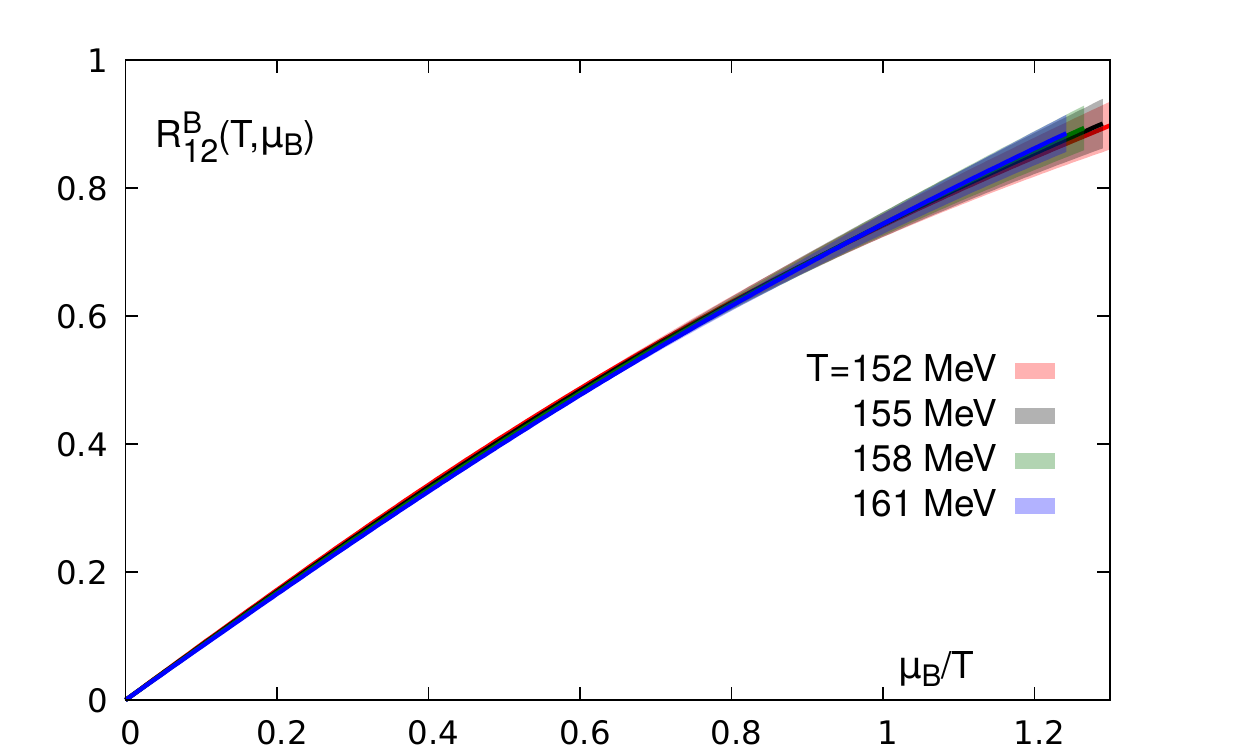}
\caption{Continuum estimate for $R_{12}^B$ based on NNNLO
        expansion results obtained on lattices
        of size $32^3\times 8$ and $48^3\times 12$. 
}
\label{fig:R12Bcont}
\end{center}
\end{figure}
The resulting continuum estimates for $R_{12}^B$, evaluated for
several values of the temperature in the vicinity of the pseudo-critical 
temperature, $T_{pc}(0)$, are shown in Fig.~\ref{fig:R12Bcont}. 
We note that the variation with temperature is small. As will be discussed
in Section IV the results obtained for $R_{12}^B$ at $\mu_B\lsim 125$~MeV  
are in good agreement with HRG model calculations.
For larger 
values of $\mu_B$ we find, however, $R_{12}^{B,QCD} > R_{12}^{B,HRG}$, which
reflects the large deviations of higher order cumulants, evaluated in QCD
at $\mu_B=0$, from the corresponding HRG values.

\subsection{Skewness and kurtosis of net baryon-number fluctuations}

While the low order cumulants $M_B=\chi_1^B(T,\mu_B)$, 
$\sigma_B^2=\chi_2^B(T,\mu_B)$ and their ratio are in good agreement with
HRG model calculations that use  non-interacting, point-like hadrons
at and below $T_{pc}$ (see also discussion in section IV), 
this clearly is not the case for higher order cumulants. This is apparent 
in calculations of the
skewness and kurtosis ratios, 
\begin{eqnarray}
	R_{31}^B(T,\mu_B) &=& \frac{S_B \sigma_B^3}{M_B} = \frac{\chi_3^B(T,\mu_B)}{\chi_1^B(T,\mu_B)} \; , \\
R_{42}^B(T,\mu_B) &=& \kappa_B \sigma_B^2 = \frac{\chi_4^B(T,\mu_B)}{\chi_2^B(T,\mu_B)} \; ,
\label{R31B}
\end{eqnarray}
which both are unity in 
non-interacting HRG model calculations, but are known to be significantly 
smaller in lattice QCD calculations already in the vicinity of the 
pseudo-critical temperature, $T_{pc}(0)$, at vanishing values of the baryon
chemical potential. Moreover, in contrast 
to the cumulant ratio $R_{12}^B$, the ratios $R_{31}^B$ and $R_{42}^B$
show a much stronger temperature dependence and a milder dependence on
$\mu_B$. It thus has been suggested that the ratio $R_{12}^B$ is well 
suited to determine the baryon chemical potential from experimental data, 
while the ratios $R_{31}^B$ and $R_{42}^B$ constrain the temperature
\cite{Karsch:2012wm,Bazavov:2012vg}.
\begin{figure}[t]
\includegraphics[width=0.47\textwidth]{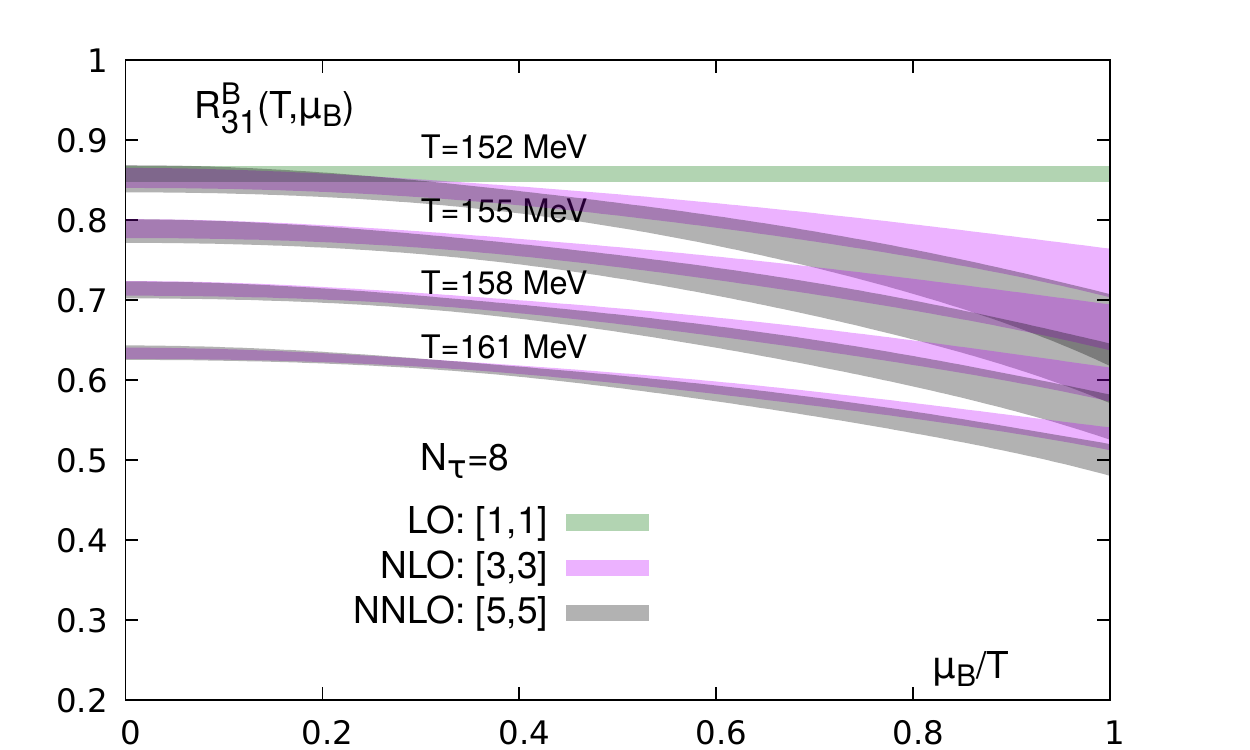}
\includegraphics[width=0.47\textwidth]{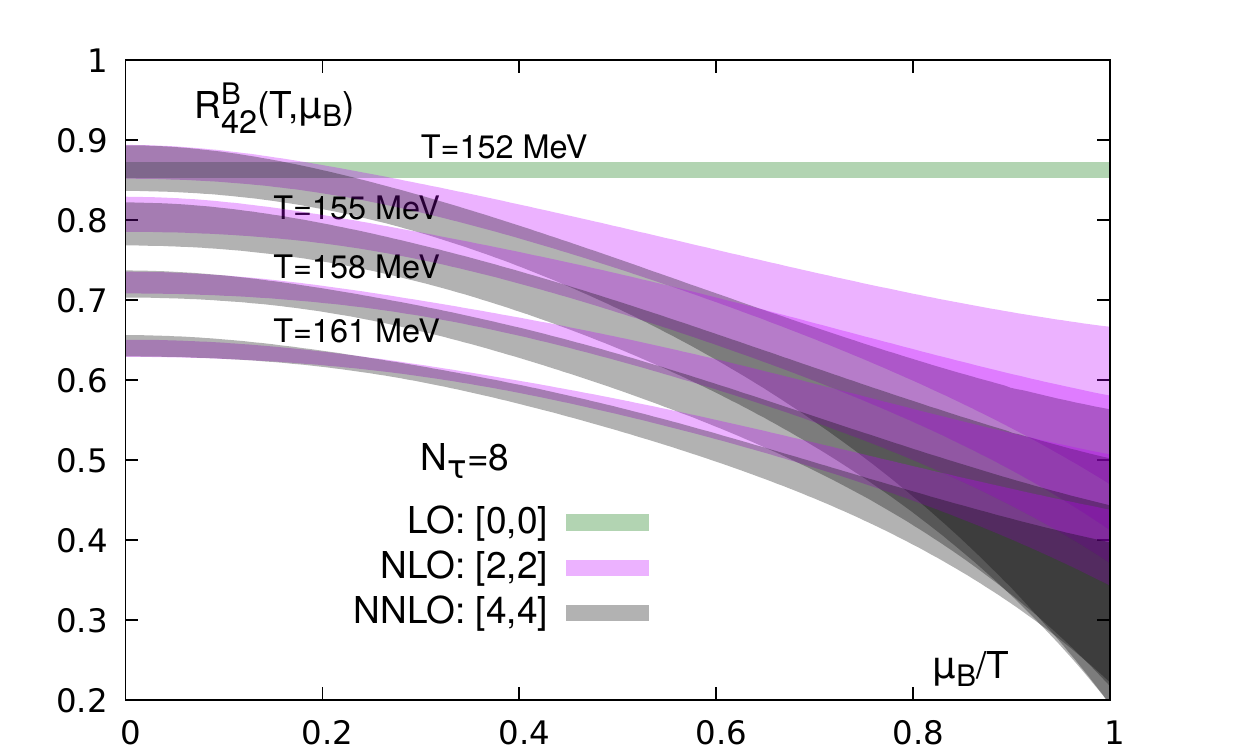}
\caption{The cumulant ratios $R_{31}^B(T,\mu_B)\equiv S_B \sigma_B^3/M_B$ (top)
and  $R_{42}^B(T,\mu_B)\equiv \kappa_B\sigma_B^2$ (bottom)
versus $\mu_B/T$ for four different values of the temperature calculated
from LO, NLO and NNLO Taylor expansions of the cumulants
$\chi_n^B(T,\mu_B)$ on lattices with temporal extent  $N_\tau=8$.
}
\label{fig:R31B}
\end{figure}

Using our results for up to $8^{th}$ order cumulants of conserved
charge fluctuations and correlations, we can construct NNLO expansions
for the third and fourth order cumulants $\chi_3^B(T,\mu_B)$ and
$\chi_4^B(T,\mu_B)$, where again the electric-charge and strangeness
chemical potentials have been fixed by demanding $n_S=0$ and $n_Q/n_B=0.4$. 
With this we determine up to NNLO results
for the skewness and kurtosis cumulant ratios $R_{31}^B$ and $R_{42}^B$.

We again first use our high statistics data obtained on the $N_\tau=8$ 
lattices to analyze the effect of truncations of the Taylor expansions
at finite orders of $\mu_B$. We used the
fit ansatz given in Eq.~\ref{fitansatz} and performed fits to LO, NLO and
NNLO results for the ratios $R_{31}^B$ and $R_{42}^B$ in the temperature
range $[135  {\rm MeV} : 175  {\rm MeV}]$ and for baryon chemical potentials
$\mu_B\le 160$~MeV. Results from these fits are shown in Fig.~\ref{fig:R31B}
for four values of the temperature in the vicinity of the pseudo-critical
temperature $T_{pc}(0)$. The two central $T$-values,
$T=155$~MeV and $158$~MeV, correspond to the lower and upper end of the 
error band for the pseudo-critical temperature at $\mu_B=0$. 
The lowest temperature, $T=152$~MeV reflects the lowest $T$-value 
reached on the pseudo-critical line $T_{pc}(\mu_B)$ at $\mu_B/T=1$.
For clarity we show in Fig.~\ref{fig:R31B} the LO results, which are 
$\mu_B$-independent, only for the lowest temperature. Of course, at all temperature
values the LO results coincide with the values of $R_{31}^B$ and $R_{42}^B$
at $\mu_B=0$. We also note that in the range of chemical potentials,
$0\le \mu_B/T\le 1$, the pseudo-critical temperature only varies 
slightly. The data 
shown in Fig.~\ref{fig:R31B} thus cover the entire parameter range
of relevance for the calculation of these cumulant ratios on the
pseudo-critical line for $\mu_B/T\lsim 1$.

\begin{figure}[t]
\begin{center}
	\includegraphics[width=85mm]{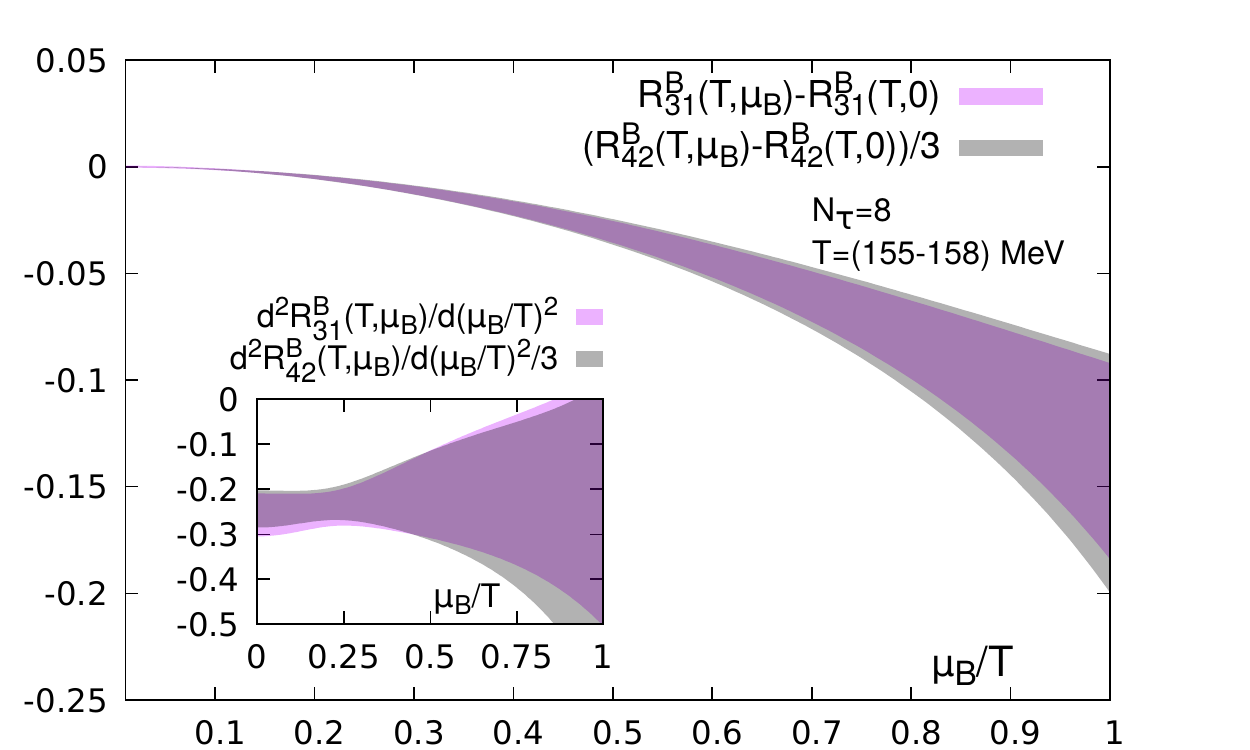}
\caption{The $\mu_B$-dependent correction to $R_{31}^B$ compared to
one third of the correction for $R_{42}^B$. The inset shows a comparison
of the second derivative of $R_{31}^B$ and $R_{42}^B/3$ with respect to
$\mu_B/T$.
}
\label{fig:slope31}
\end{center}
\end{figure}

In \cite{Bazavov:2017tot} we showed that the skewness and kurtosis ratios 
$R_{31}^B$ and $R_{42}^B$ are almost identical in leading order, 
${\cal O}(\mu_B^0)$. The NLO correction to the kurtosis ratio $R_{42}^B$, 
however, is about a factor three larger than that for the 
skewness ratio $R_{31}^B$. Fig.~\ref{fig:R31B} suggests 
that these relations are still well respected by the NNLO results. 
The slope of $R_{42}^B(T,\mu_B)$ as function of $\hat{\mu}_B$ at fixed $T$ 
is significantly larger than that of 
$R_{31}^B(T,\mu_B)$ and, in fact, it is still consistent with being about
a factor three larger. This is shown in
Fig.~\ref{fig:slope31} where we compare the $\mu_B$-dependent parts
of $R_{31}^B$ and $R_{42}^B/3$. Also shown in this figure are
the second derivatives of $R_{31}^B(T,\mu_B)$  and $R_{42}^B(T,\mu_B)/3$ with 
respect to $\mu_B/T$.

Compared to the lower order ratio $R_{12}^B$ higher order corrections
in the Taylor expansion of $R_{31}^B$ are significantly larger. 
In the temperature range shown in Fig.~\ref{fig:R31B}
corrections to the NLO results, arising from the NNLO, ${\cal O}(\mu_B^5)$,
corrections in the Taylor expansions of the cumulants 
$\chi_3^B(T,\mu_B)$, are about 5\% at
$\mu_B/T = 0.8$ and rise to about 10\% at $\mu_B/T = 1$. Consequently
truncation effects in  $R_{42}^B$ are about a factor three larger.

\begin{figure}[t]
\includegraphics[width=0.47\textwidth]{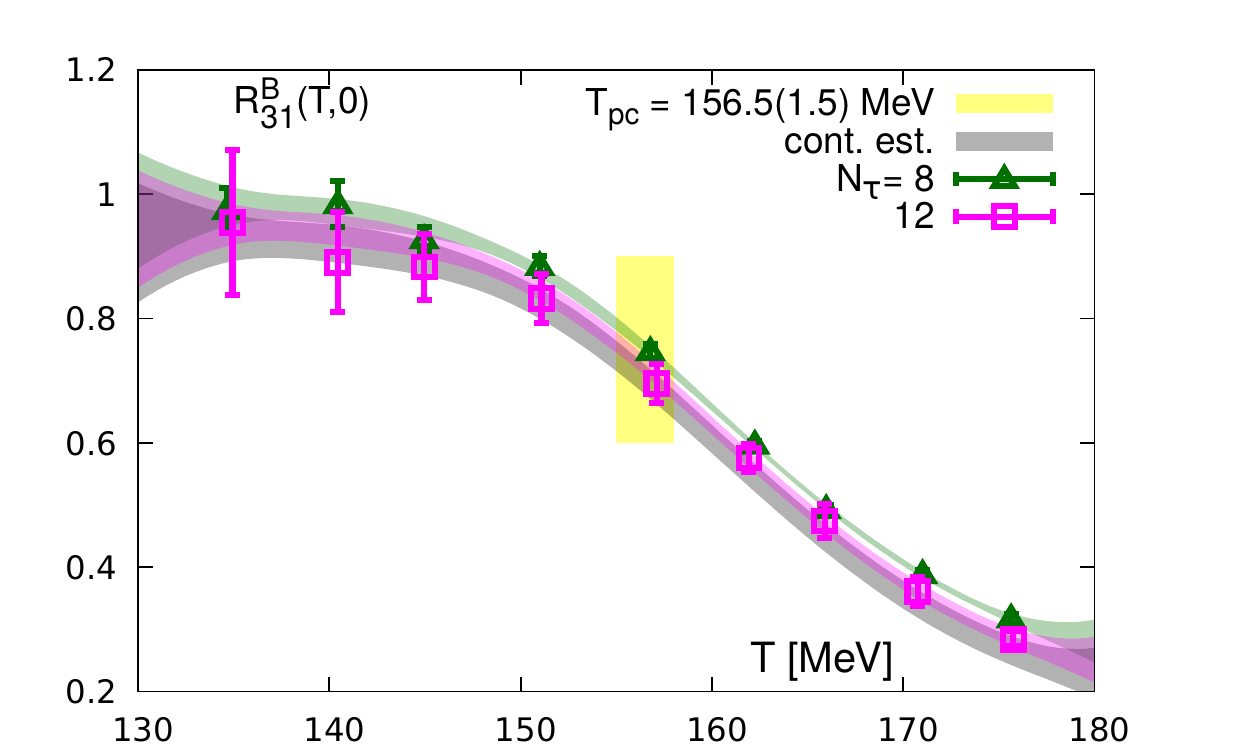}
\includegraphics[width=0.47\textwidth]{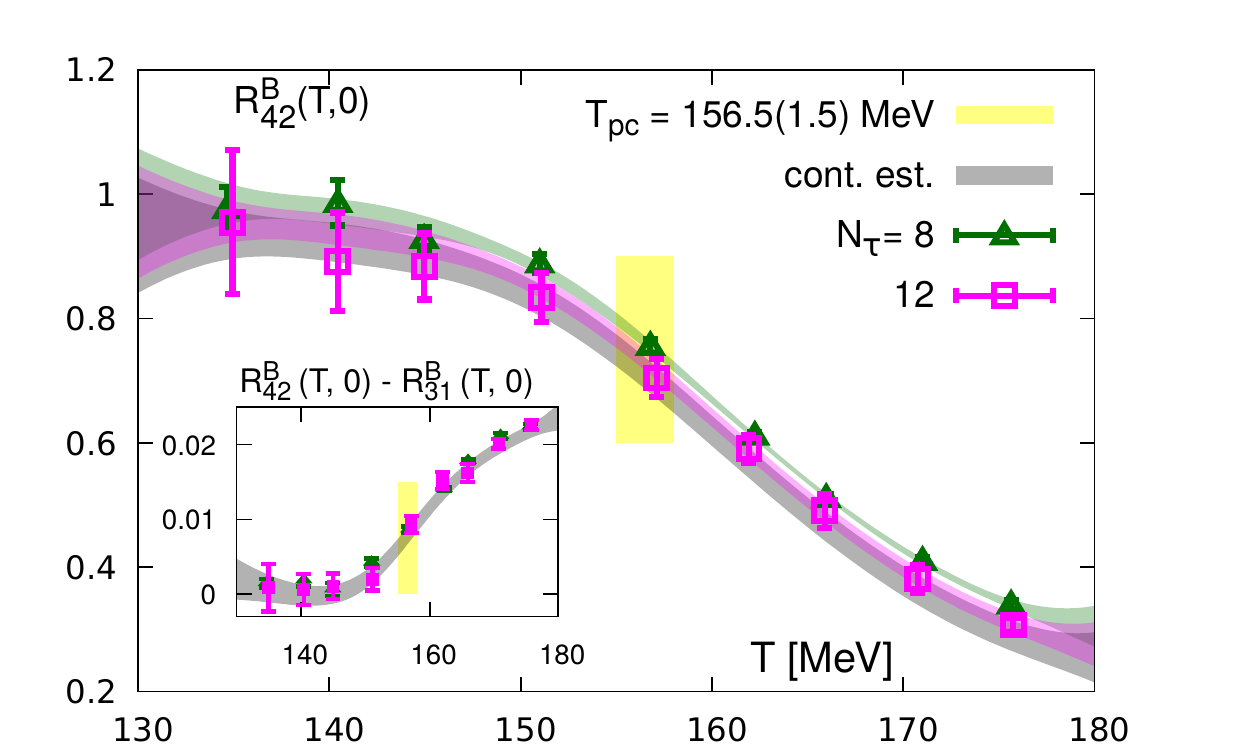}
\caption{Continuum estimates for the skewness ratio, 
$R_{31}^B\equiv S_B\sigma_B^3/M_B$ (top), and kurtosis ratio 
$R_{42}^B\equiv \kappa_B\sigma_B^2$ (bottom) at $\mu_B=0$
based on results obtained on lattices of size $32^3\times 8$ 
and $48^3\times 12$, respectively. The inset in the bottom figure
shows the difference $R_{42}^B-R_{31}^B$ at $\mu_B=0$ as function
of $T$.
}
\label{fig:Rnm-nt8nt12}
\end{figure}

In Fig.~\ref{fig:Rnm-nt8nt12} we show results for the skewness
and kurtosis ratios $R_{31}^B(T,\mu_B)$ and $R_{42}^B(T,\mu_B)$
obtained at $\mu_B=0$ on lattices with temporal extent $N_\tau=8$ and $12$. 
Obviously results for $N_\tau=12$ are systematically
below those for $N_\tau=8$. This is in accordance with the observed
shift of the pseudo-critical temperatures \cite{Bazavov:2018mes} to 
smaller values with increasing $N_\tau$ or, equivalently,
decreasing lattice spacing at fixed temperature $aT =1/N_\tau$.
When performing joint fits to the $N_\tau=8$ and $12$ data, using
the ansatz given in Eq.~\ref{joint-fit}, we find that within our 
current statistical errors on the $N_\tau=12$ we cannot resolve 
any $T$- or $\mu_B/T$-dependence of cut-off effects. It thus suffices 
to use a constant ansatz for the cut-off corrections, {\it i.e.} we
use $g(T,\mu_B)= a_{0,0}$  and a $[3,4]$ rational polynomial for
the continuum limit result $f(T,\mu_B)$. A joint fit 
to the $N_\tau=8$ and $12$ data yields $a_{0,0}=3.2(1.5)$ for 
$R_{31}^B(T,\mu_B)$  and $a_{0,0}=3.2(3.0)$ for $R_{42}^B(T,\mu_B)$.
The resulting continuum limit 
estimates at $\mu_B=0$ are also shown in Fig.~\ref{fig:Rnm-nt8nt12}. 

The inset in Fig.~\ref{fig:Rnm-nt8nt12}~(bottom)
shows the continuum estimate for the difference 
$R_{42}^B-R_{31}^B$ at $\mu_B=0$ as function of $T$. At temperatures
below $T\simeq 150$~MeV this difference is consistent with being
zero. In the crossover region, $T_{pc}(0)=156.5(1.5)$~MeV we find
that the difference is slightly positive, 
$R_{42}^B (T_{pc})-R_{31}^B(T_{pc})=0.008(3)$.

\begin{figure}[htb]
\includegraphics[width=0.47\textwidth]{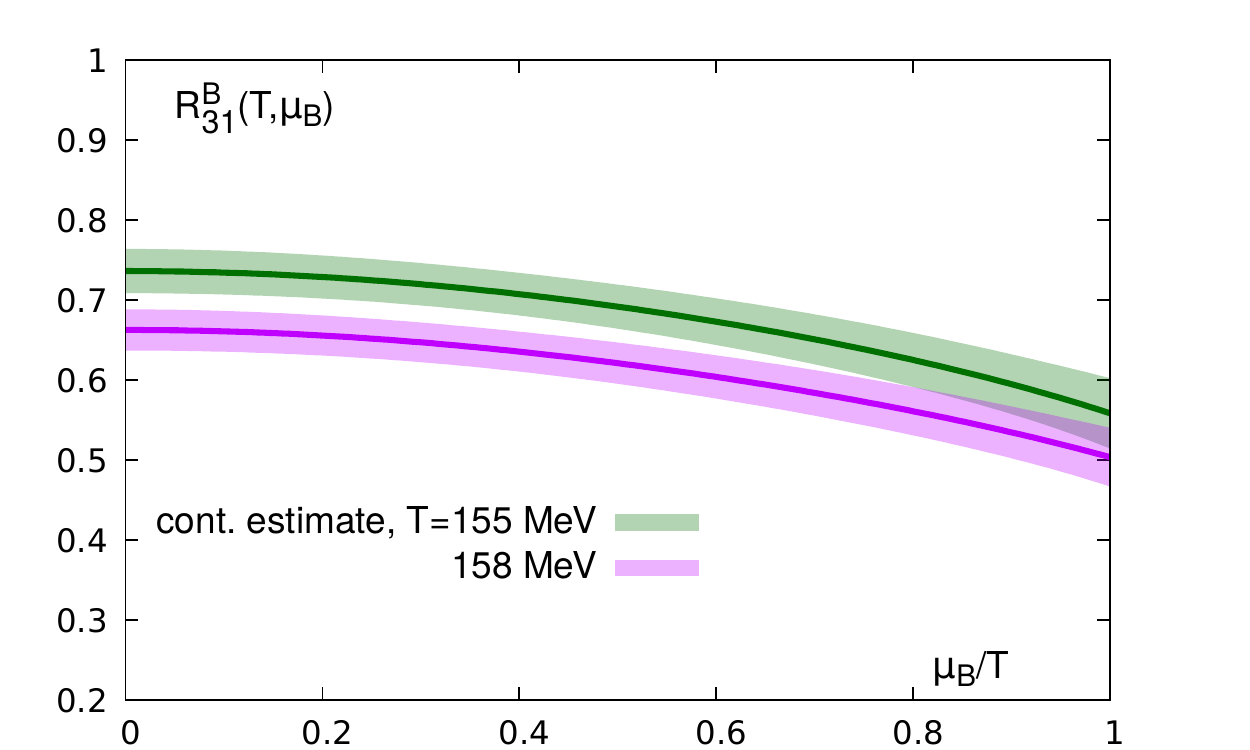}
\includegraphics[width=0.47\textwidth]{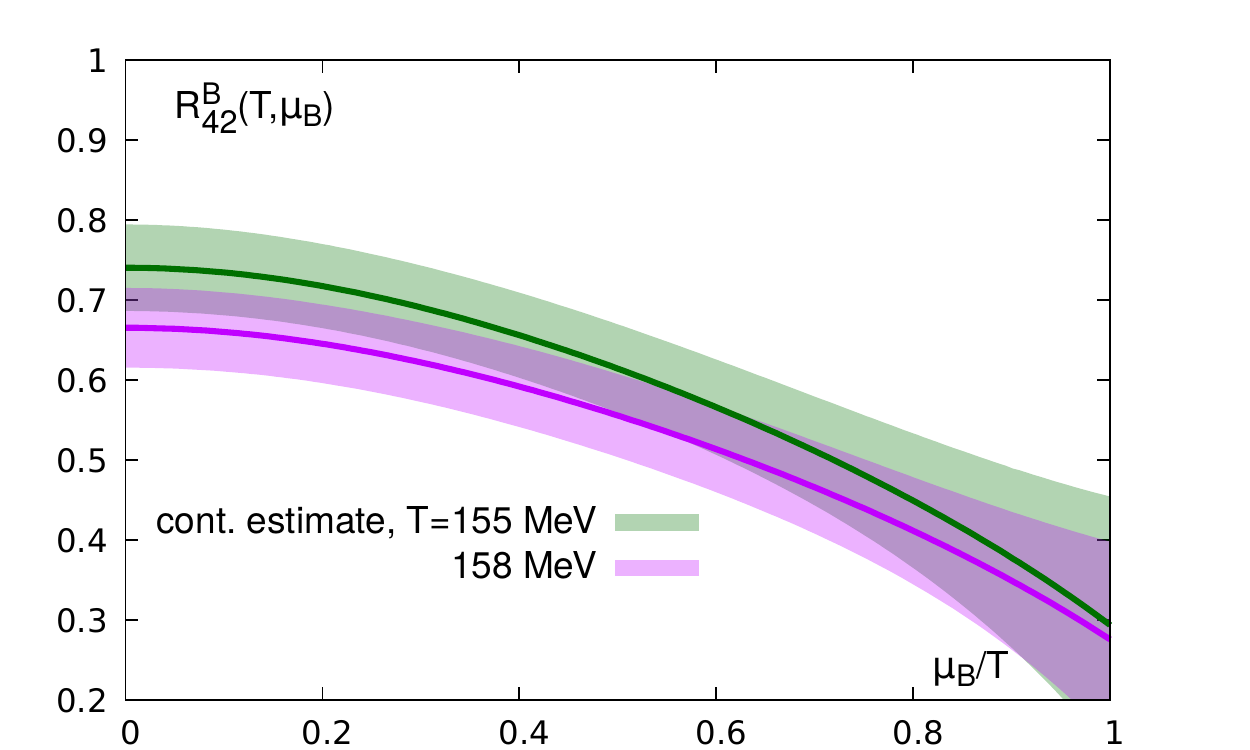}
\caption{Continuum estimates for the skewness (top) and kurtosis (bottom) 
ratios obtained from joint fits to data obtained on lattices with temporal 
extent $N_\tau=8$ and $12$.}
\label{fig:continuum}
\end{figure}

Continuum estimates for $R_{31}^B(T,\mu_B)$ and  $R_{42}^B(T,\mu_B)$ at 
two values of the temperature, corresponding to the current error band
for the pseudo-critical temperature at $\mu_B=0$ are shown in
Fig.~\ref{fig:continuum}. 

\subsection{Hyper-skewness and hyper-kurtosis of net baryon-number fluctuations}

The $5^{th}$ and $6^{th}$ order cumulants are related to the corresponding
$5^{th}$ and $6^{th}$ order standardized moments, {\it i.e.} 
the hyper-skewness, $S^H$, and hyper-kurtosis, $\kappa^H$. We consider
here the cumulant ratios for $5^{th}$ and $6^{th}$ order cumulants of net
baryon-number fluctuations,
\begin{eqnarray}
	R_{51}^B(T,\mu_B) \equiv \frac{S^H_B\sigma_B^5}{M_B} 
	&=&\frac{\chi_5^B(T,\mu_B)}{\chi_1^B(T,\mu_B)} \; ,
\nonumber \\
R_{62}^B(T,\mu_B) \equiv \kappa^H_B \sigma_B^4 
&=&\frac{\chi_6^B(T,\mu_B)}{\chi_2^B(T,\mu_B)} \; .
\label{hyper}
\end{eqnarray}

Unlike the ratios for skewness and kurtosis cumulants,
the corresponding ratios involving $5^{th}$ and $6^{th}$ order cumulants 
are negative already at $\mu_B=0$ in a broad temperature interval in the 
vicinity of $T_{pc}(0)$ and become smaller with increasing $\mu_B$. This
reflects the properties of the $6^{th}$ and  $8^{th}$ order cumulants
shown in Fig.~\ref{fig:Bdiag}.

The $\mu_B$-dependence of the cumulant ratios
$R_{51}^B$ and $R_{62}^B$ follows a pattern similar to that of the 
skewness and kurtosis ratios. In particular, in LO both ratios are almost 
identical and the NLO correction to $R_{62}^B$ is about a factor three larger
than that for $R_{51}^B$.
Like in the case of corresponding relations
for the skewness and kurtosis ratios these relations simply result
from the structure of Taylor expansions for odd and even cumulants
\cite{Bazavov:2017tot}. The
relations are exact for expansions at vanishing $\mu_Q$ and $\mu_S$ and
apparently they are not much altered in the strangeness neutral
case $n_S=0$ with $n_Q/n_B=0.4$. A fit to the $N_\tau=8$ lattice QCD results
for the difference $R_{62}^B - R_{51}^B$ at $\mu_B=0$ yields $0.029(9)$.

\begin{figure}[t]
\includegraphics[width=0.47\textwidth]{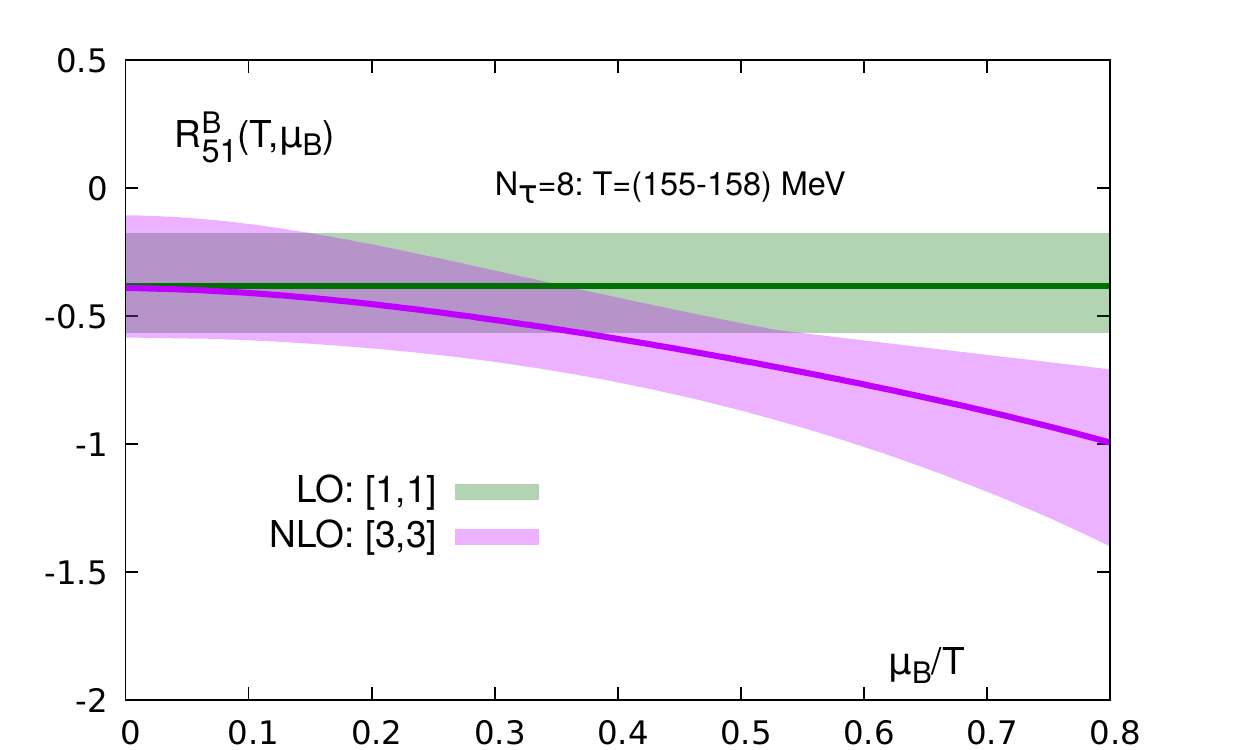}
\includegraphics[width=0.47\textwidth]{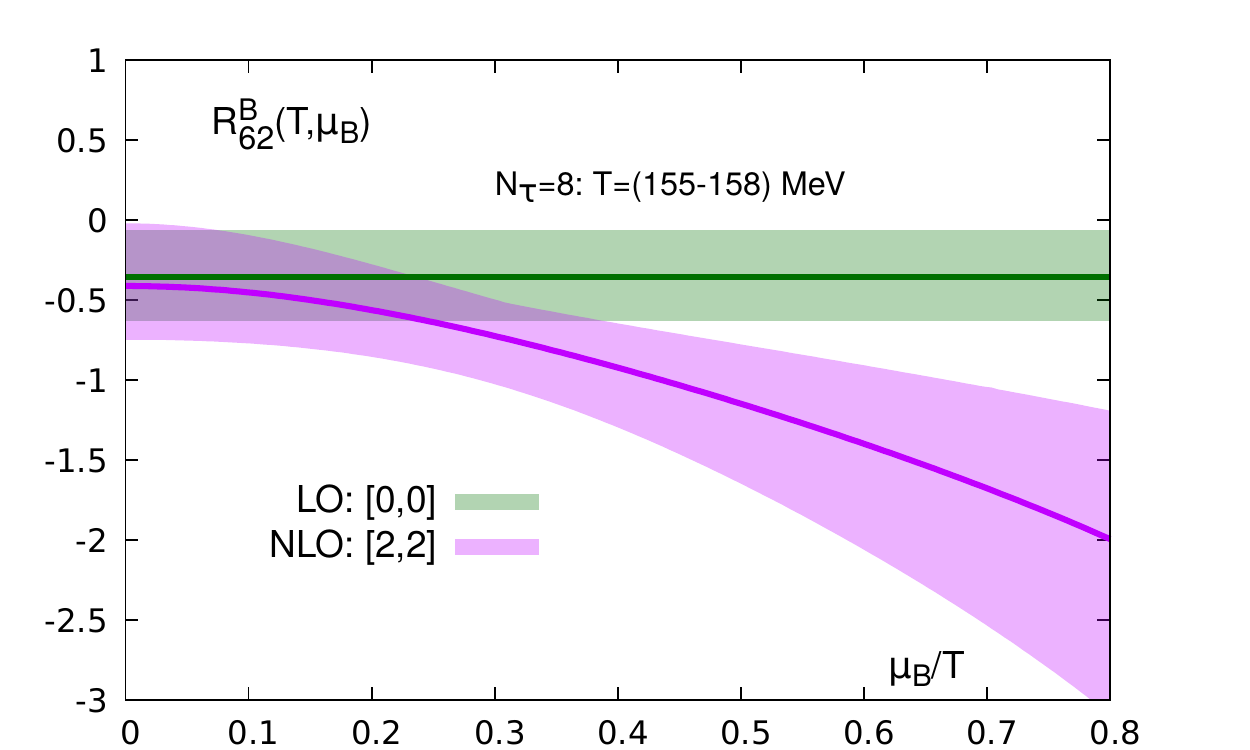}
\caption{The cumulant ratios $R_{51}^B(T,\mu_B)$ and $R_{62}^B(T,\mu_B)$
versus $\mu_B/T$ 
from LO and NLO Taylor expansions of the cumulants calculated on
lattices with temporal extent  $N_\tau=8$.
}
\label{fig:R51B}
\end{figure}

While statistical errors are strongly correlated between the $5^{th}$ and 
$6^{th}$ order cumulants they are large for each of these 
cumulants individually. For this reason we only present results
for these cumulants obtained on lattices with temporal extent $N_\tau=8$
and evaluate the NLO corrections only for $\mu_B/T\le 0.8$.
NLO results for $R_{51}^B(T,\mu_B)$ and $R_{62}^B(T,\mu_B)$ are shown in 
Fig.~\ref{fig:R51B}. 

Obviously NLO corrections for these ratios are negative and substantially 
larger than those in the skewness and kurtosis ratios.
In the vicinity of the pseudo-critical temperature the difference between 
LO and NLO results at $\mu_B/T=0.8$  is about an order of magnitude larger 
in $R_{51}^B(T,\mu_B)$ than in $R_{31}^B(T,\mu_B)$. This also is the
case when comparing  $R_{62}^B(T,\mu_B)$ with  $R_{42}^B(T,\mu_B)$.

The magnitude and sign of the NLO corrections to $5^{th}$ and $6^{th}$
order cumulants in relation to corresponding results for the $3^{rd}$ and 
$4^{th}$ order cumulants is evident from the structure of the corresponding
Taylor expansion coefficients. It is easy to see this in Taylor
expansions performed at $\mu_Q=\mu_S=0$. In this case one has, for instance,
\begin{eqnarray}
	\chi_4^B(T,\mu_B) &=& \chi_4^B+ \frac{\chi_6^B}{2}
	\left( \frac{\mu_B}{T}\right)^2 +\frac{\chi_8^B}{24}\left( \frac{\mu_B}{T}\right)^4 + ..
\; , \\
\chi_6^B(T,\mu_B) &=& \chi_6^B+ \frac{\chi_8^B}{2}
\left( \frac{\mu_B}{T}\right)^2 + ...
\label{simple}
\end{eqnarray}
As can be deduced from Fig.~\ref{fig:Bdiag}, despite of the large errors on 
current results for $\chi_8^B$, the cumulants $\chi_6^B$ and 
$\chi_8^B$ are both negative in the vicinity of the
pseudo-critical temperature, however the absolute value of the $8^{th}$ order
cumulant is about an order of magnitude larger. This results in the
much larger NLO correction to the expansion of $\chi_6^B(T,\mu_B)$.
Although the expansions of all cumulants $\chi_n^B(T,\mu_B)$ will have 
the same radius of convergence it is apparent that expansions for 
higher order cumulants will converge more slowly. Higher order corrections 
to $\chi_5^B(T,\mu_B)$ and $\chi_6^B(T,\mu_B)$ will thus be needed to arrive
at firm conclusions on the behavior of these cumulants close to
$\mu_B/T \simeq 1$. For $\mu_B/T\simeq 0.3$, however, the NLO correction
is about an order of magnitude smaller and thus of similar magnitude as
the NNLO correction to $\chi_3^B(T,\mu_B)$ and $\chi_4^B(T,\mu_B)$
at $\mu_B/T \simeq 1$.

For small values of the baryon chemical potential and $\mu_S=\mu_Q=0$
we thus may extent the result on the ordering of cumulant ratios stated in 
Eq.~\ref{ordering} and
include also results for the $5^{th}$ and $6^{th}$ order cumulant ratios,
\begin{eqnarray}
	\frac{\chi_6^B(T,\vec{\mu})}{\chi_2^B(T,\vec{\mu})} <
	\frac{\chi_5^B(T,\vec{\mu})}{\chi_1^B(T,\vec{\mu})} <
\frac{\chi_4^B(T,\vec{\mu})}{\chi_2^B(T,\vec{\mu})}  <
        ~\frac{\chi_3^B(T,\vec{\mu})}{\chi_1^B(T,\vec{\mu})} \; .
        \label{ordering3456}
\end{eqnarray}

\section{Baryon-number fluctuations on the pseudo-critical line 
and the cumulants of net proton-number fluctuations}

In this section we compare results on higher order cumulants of net 
proton-number fluctuations, obtained by the STAR Collaboration during BES-I 
at RHIC \cite{Adam:2020unf,STAR54}, with our results for cumulants of net 
baryon-number fluctuations calculated in QCD on the pseudo-critical line given 
in Eq.~\ref{Tpc}.
The pseudo-critical line shows only a rather weak dependence on $\mu_B$.
The ${\cal O}(\mu_B^4)$ correction to $T_{pc}(\mu_B)$ is found to be
zero within errors \cite{Bazavov:2017tot}.
For $\mu_B\le T_{pc}(0)$ it changes from $T=156.5(1.5)$~MeV to 
$154.5(2.0)$~MeV. This range of temperatures is well covered by the
results for cumulant ratios as function of $\mu_B$ 
evaluated at fixed values of the temperature that have been
shown in the previous section.

\begin{figure}[t]
\includegraphics[width=0.47\textwidth]{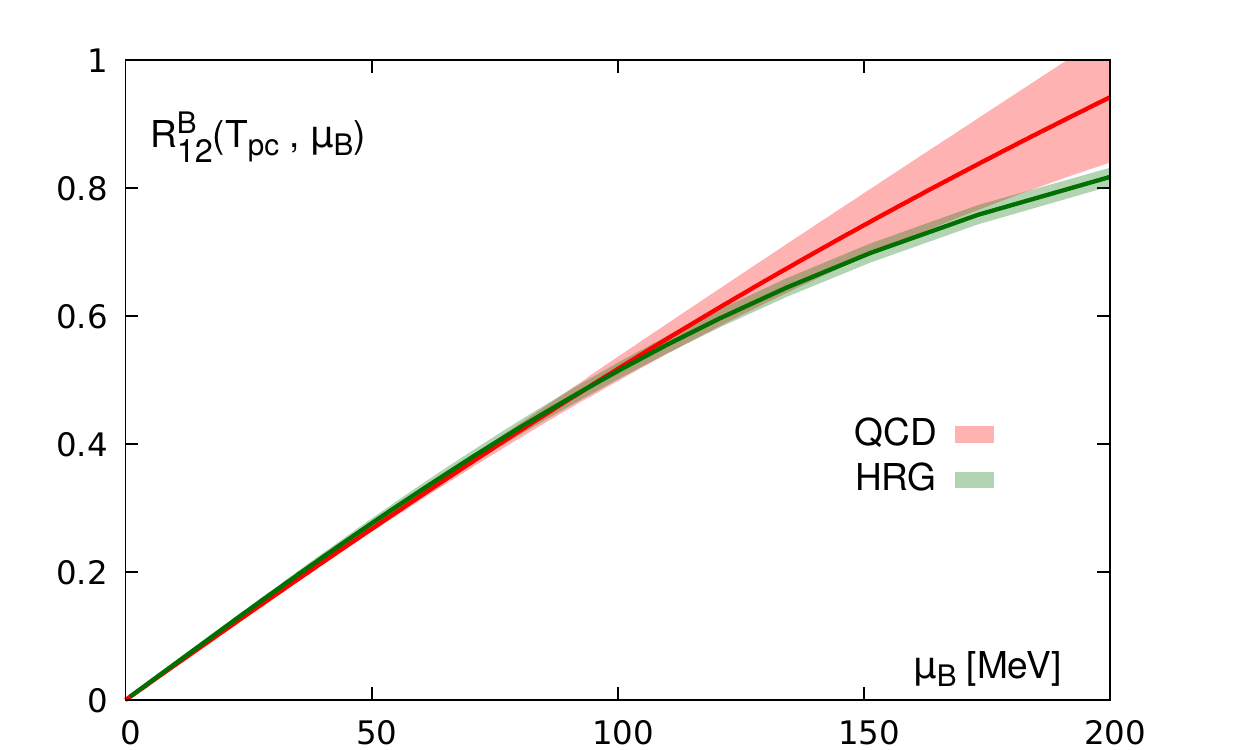}
\caption{
The cumulant ratio $R_{12}^B(T,\mu_B)$ evaluated on the pseudo-critical
line $T_{pc}(\mu_B)$ for the case $n_S=0$ and $n_Q/n_B=0.4$. Also shown
is the corresponding result obtained in HRG model calculations.
In the latter case the
width of the line reflects differences resulting from using
particle spectra for a non-interacting HRG listed in the Particle Data
Tables as well as resulting within Quark Model calculations.
}
\label{fig:pseudo}
\end{figure}

In Fig.~\ref{fig:pseudo} we show results for $R_{12}^B(T_{pc}(\mu_B),\mu_B)$ 
on the pseudo-critical line and compare with results obtained from 
non-interacting HRG model calculations that utilize hadron resonance
gas spectra as listed in the particle data tables \cite{PDG} as well 
as spectra calculated in Quark Models \cite{Isgur,Ebert}. 
As can be seen in Fig.~\ref{fig:pseudo} HRG model calculations for 
$R_{12}^B$ agree well with QCD results obtained on the pseudo-critical
line up to about $\mu_B/T\simeq 0.8$ or $\mu_B \simeq 125$~MeV. This suggests
that the use of low order HRG cumulants, in particular the mean of
hadron distributions (hadron yields) that are used experimentally to determine 
freeze-out parameters, may be appropriate at small values of the baryon chemical
potential or small net baryon-number densities. The HRG model estimates
of freeze-out parameters \cite{Adamczyk:2017iwn} suggest that the range of 
baryon chemical potentials
$\mu_B/T\lsim 1$ corresponds to thermal conditions at freeze-out generated 
in heavy ion experiments at beam energies $\sqrt{s_{_{NN}}}\gsim 27$~GeV.
Fig.~\ref{fig:pseudo} suggests that below this value of $\sqrt{s_{_{NN}}}$
HRG model determinations of baryon chemical potentials differ from QCD
determinations by more than 10\%. It thus may be useful to eliminate
$\mu_B$ in favor of a directly accessible physical observable, e.g.
$R_{12}^B$.

\begin{figure}[t]
\includegraphics[width=0.47\textwidth]{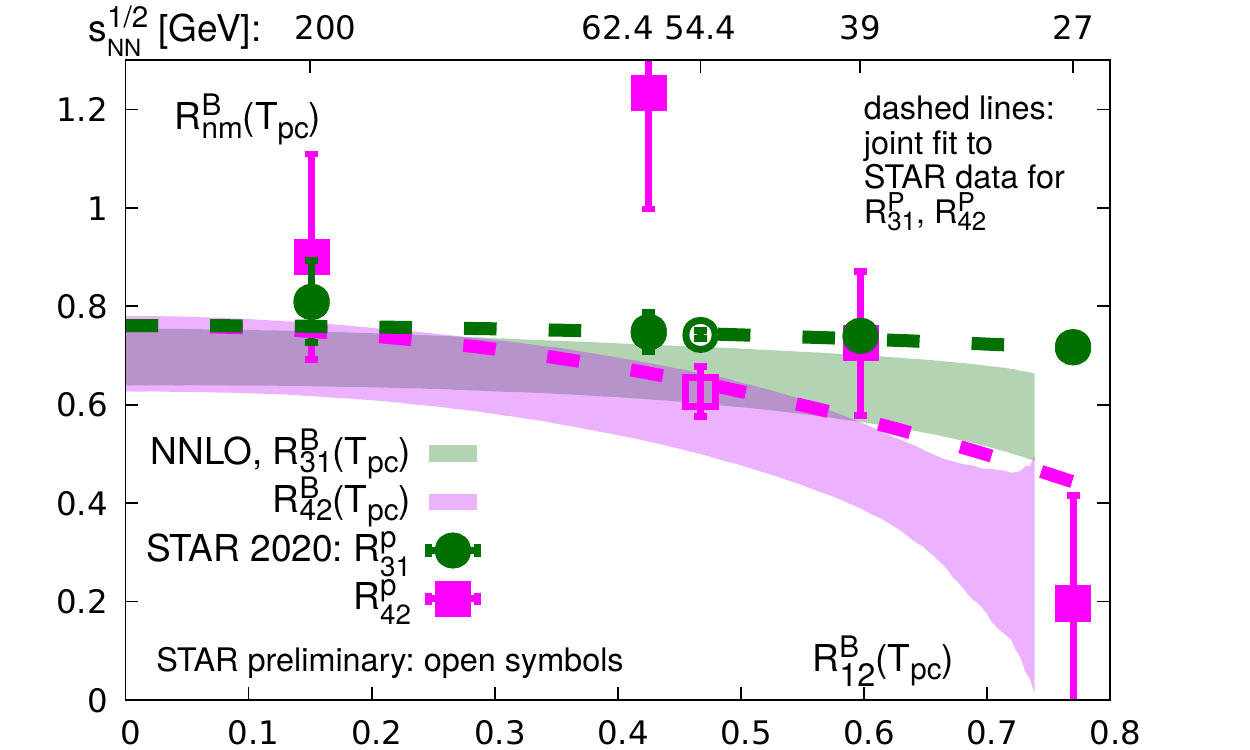}
\caption{The cumulant ratios (bands)
$R_{31}^B(T,\mu_B)\equiv S_B \sigma_B^3/M_B$ and
$R_{42}^B(T,\mu_B)\equiv \kappa_B \sigma_B^2$ versus
$R_{12}^B(T,\mu_B) \equiv M_B/\sigma_B^2$
on the pseudo-critical line calculated from NNLO Taylor series.
Data are results on cumulant ratios of net proton-number
fluctuations obtained by the STAR Collaboration \cite{Adam:2020unf}.
Also shown are preliminary results obtained at
$\sqrt{s_{_{NN}}}=54.4$~GeV \cite{STAR54}.
Dashed lines show joint fits to the data as described in the text.
}
\label{fig:pseudo42}
\end{figure}

\begin{figure}[t]
\includegraphics[width=0.47\textwidth]{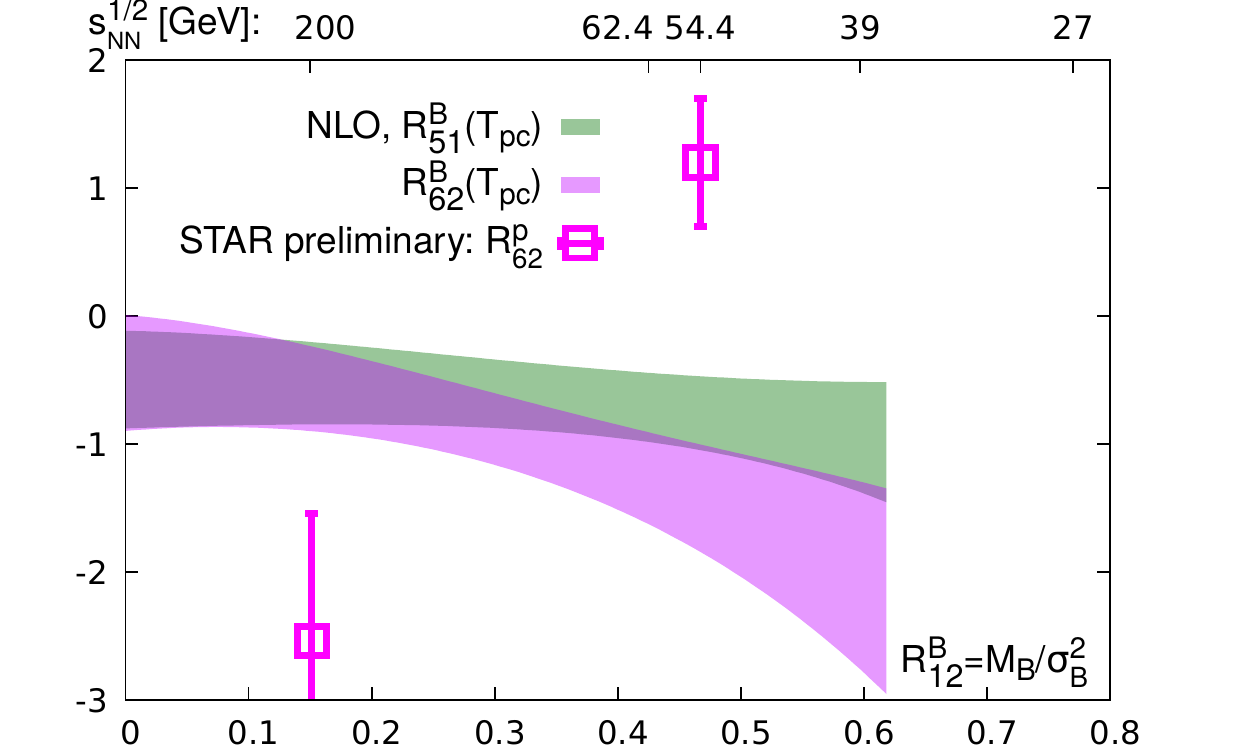}
\caption{
The cumulant ratios $R_{51}^B(T,\mu_B)$  and $R_{62}^B(T,\mu_B)$ vs.
$R_{12}^B(T,\mu_B)$ evaluated on the pseudo-critical line.
Data are preliminary results for the cumulant ratio $R_{62}^P$ of
net proton-number fluctuations obtained by the STAR Collaboration
at $\sqrt{s_{_{NN}}}=200$~GeV and $54.4$~GeV for the (0-40)\% centrality
class \cite{STAR54}.
}
\label{fig:pseudo62}
\end{figure}

At least for $\mu_B \lsim 200$~MeV truncation errors in the 
Taylor expansion of the first two cumulants, mean and variance, as well as 
lattice discretization errors are small. The continuum limit extrapolation for 
$R_{12}^B(T_{pc}(\mu_B),\mu_B)$, shown in Fig.~\ref{fig:pseudo} 
thus does not suffer from truncation errors in the Taylor series
at least up to $\mu_B/T=1.2$. It is a 
monotonically rising function\footnote{Note that this will no longer be the	
case when one comes close to a critical point, where $\chi_2^B$ is expected to 
diverge and $R_{12}^B(T_{pc}(\mu_B),\mu_B)$ thus would approach zero.}
of $\mu_B$. This allows to replace the chemical potential in an analysis of 
higher order cumulant ratios in favor of $R_{12}^B$. We have done so for the
comparison of higher order cumulant ratios calculated in lattice QCD 
on the pseudo-critical line with experimental data on cumulants
of net proton-number fluctuations. In Fig.~\ref{fig:pseudo42} we show the
skewness and kurtosis ratios, $R_{31}^B$ and $R_{42}^B$,
on the pseudo-critical line as function
of $R_{12}^B$, which also has been evaluated on the pseudo-critical line. 
Similar results for the hyper-skewness and hyper-kurtosis ratios are shown
in Fig.~\ref{fig:pseudo62}.

In Fig.~\ref{fig:pseudo42} we show lattice QCD results up to $R_{12}^B=0.75$,
which corresponds to $\mu_B=T_{pc}(\mu_B)\simeq 154.5$~MeV. The width of the 
bands shown in the figure reflect 
the error on $T_{pc}(\mu_B)$ as given in Eq.~\ref{Tpc} as well as the error
on the NNLO and continuum limit estimates for $R_{31}^B$ and
$R_{42}^B$. Note that the upper end of these error bands correspond to the 
lower temperature, {\it i.e.} $T=155$~MeV at $\mu_B=0$ and 
$T\simeq 152.5$~MeV at $\mu_B/T=1$.

Also shown in this figure are results for the skewness and kurtosis ratios 
of net proton-number fluctuations obtained by the STAR Collaboration
\cite{Adam:2020unf,STAR54}. These 
ratios are plotted versus the measured ratio of mean over variance of net 
proton-number fluctuations, which is taken as a proxy for the net 
baryon-number cumulant 
ratio\footnote{In a non-interacting HRG with vanishing
        strangeness and electric-charge chemical potential the mean over
        variance of net proton-number fluctuations and net baryon-number
        fluctuations are identical. In the case of a strangeness neutral 
        ($n_S=0$ with $n_Q/n_B=0.4$), non-interacting HRG, however, the 
	latter is about 10\% smaller.}  $R_{12}^B$.

As the experimentally determined skewness ratio of net proton-number
fluctuations has a rather weak dependence 
on $R_{12}^P$ and also the QCD result for $R_{31}^B$ has a weak 
dependence on $R_{12}^B$, it obviously is not of much importance for
the comparison of data and lattice QCD calculations whether
$R_{12}^P$ equals $R_{12}^B$ or only is a proxy within say (10-20)\%.
More relevant is the question
to what extent the magnitude of $R_{31}^P$
is a good approximation\footnote{ 
Many caveats for a direct comparison between net baryon-number fluctuations
calculated in equilibrium thermodynamics and net proton-number fluctuations
measured in heavy ion collisions have been discussed in the literature 
\cite{Bzdak:2019pkr,Ratti}. 
The lattice QCD results shown in Fig.~\ref{fig:pseudo42} thus may be
considered only as a starting point for a more refined analysis of the
experimental data that may take into account effects arising from 
experimental acceptance cuts, the small size of the hot and dense
medium, non-equilibrium effects etc.}
for $R_{31}^B$. A direct comparison between $R_{31}^P$ and $R_{31}^B$,
as shown in  Fig.~\ref{fig:pseudo42}, suggests that freeze-out happens
in the vicinity but below the pseudo-critical temperature.  
In fact, as can be seen in Figs.~\ref{fig:R31B} and \ref{fig:continuum},
the ratios $R_{31}^B$ and  $R_{42}^B$ are decreasing functions of the
temperature. Experimental data for $R_{31}^P$ lying above the
theoretical band for $R_{31}^B$, evaluated on the pseudo-critical
line, thus suggest freeze-out to happen at a lower temperature.

Although errors on experimental results
for the kurtosis ratio $R_{42}^P$ are large, they are thermodynamically
consistent with the data on the skewness ratio as pointed out already in
our earlier analysis \cite{Bazavov:2017tot}. This gets further support
through recent high statistics\footnote{Statistics at 
$\sqrt{s_{_{NN}}}=54.4$~GeV is a factor 3.4 larger than at 
$\sqrt{s_{_{NN}}}=200$~GeV and a factor (17-30) larger than at the other
$\sqrt{s_{_{NN}}}$ data sets shown in Fig.~\ref{fig:pseudo42}.}
data obtained by
the STAR Collaboration at $\sqrt{s_{_{NN}}}=54.4$~GeV \cite{STAR54}. 
These data are shown
in Fig.~\ref{fig:pseudo42} at $R_{12}^P=0.4672(2)$. For this value of the
beam energy the kurtosis ratio $R_{42}^P$ is found to be smaller than
$R_{31}^P$. The magnitude of this difference, $R_{42}^P-R_{31}^P=-0.12(5)$, 
is in good agreement with the corresponding lattice QCD result 
on the pseudo-critical line. For the range $R_{12}^B=0.45(5)$, which 
corresponds to $\mu_B= (80-100)$~MeV, or $\mu_B/T = 0.57(7)$, we find
from a fit to the difference of $R_{42}^B$ and $R_{31}^B$,
$R_{42}^B-R_{31}^B = -0.08(3)$. At these values of the baryon chemical
potential (or for $R_{12}^B \simeq 0.5$) the NNLO results
for the skewness and kurtosis ratios, presented in the previous section,
seem to suffer little from truncation effects in the Taylor expansions.

Also shown in Fig.~\ref{fig:pseudo42} with dashed lines is a joint fit
to the experimental data on $R_{31}^P$ and $R_{42}^P$ \cite{Adam:2020unf} for
$\sqrt{s_{_{NN}}}\ge 19.6$~{\rm GeV} using a quadratic
ansatz, already used in Ref.~\cite{Bazavov:2017tot},
\begin{eqnarray}
	R_{31}^P &=& S_0  + S_2 \left(  R_{12}^P\right)^2\; , \nonumber \\
	R_{42}^P &=& K_{0} +K_{2} (R_{12}^P)^2\; ,
	\label{fitdata}
\end{eqnarray}
with $K_{0}\equiv S_{0}$. Including the new data at $\sqrt{s_{_{NN}}}= 54.4$~{\rm GeV} yields a fit, consistent with \cite{Bazavov:2017tot}, but further
constrains the parameters. One finds $S_0\equiv K_0 = 0.761(20)$ , 
$S_2=-0.077(70)$, $K_2=-0.54(22)$. From the continuum estimates of
$R_{31}^B$ and $R_{42}^B$ at $\mu_B=0$ shown in Fig.~\ref{fig:Rnm-nt8nt12}
one finds that the value of $S_0$ corresponds to a freeze-out temperature
of $153.5(2.0)$~MeV. This temperature range is consistent with an
earlier determination of the freeze-out temperature that was based
on a comparison of the mean over variance ratio of net electric-charge and
net proton-number ratios obtained by the STAR and PHENIX Collaborations
\cite{Adamczyk:2014fia,Adare:2015aqk}
with corresponding lattice QCD calculations for net electric-charge and
net baryon-number cumulant ratios \cite{Bazavov:2015zja}. We also note that 
the ratio of the curvature of $R_{42}^B$ and $R_{31}^B$ on the pseudo-critical
(freeze-out) line tends to be larger than 3, which also has been
noted in our previous analysis of skewness and kurtosis ratios 
\cite{Bazavov:2017tot}.

While the experimental data on the skewness and kurtosis cumulant ratios
of net proton-number fluctuations,
obtained at $\sqrt{s_{_{NN}}}\ge 27$~GeV, are consistent with results on
net baryon-number cumulants calculated within equilibrium QCD thermodynamics,
this is not the case for the preliminary data on $6^{th}$ order cumulants 
presented by the STAR Collaboration \cite{STAR54}. The still preliminary data 
at $\sqrt{s_{_{NN}}}= 200$~GeV and $54.4$~GeV, taken from the (0-40)\%
centrality class, are shown in Fig.~\ref{fig:pseudo62}
together with the NLO lattice QCD calculations. 
At both values of $\sqrt{s_{_{NN}}}$ deviations from the NLO lattice QCD 
results are large and of similar magnitude. While it is conceivable that
the NLO results at $R_{12}^B\simeq 0.5$ (or $\mu_B/T\simeq 0.6$) will
receive seizable corrections in NNLO, this is not the case at 
$R_{12}^B\simeq 0.15$ (or $\mu_B/T\simeq 0.3$). It thus seems impossible 
to describe both data points within QCD equilibrium thermodynamics. 
We also note that a large positive $\chi_{10}^B$ would be needed, if such
a contribution should render the hyper-kurtosis ratio to become positive 
at $\sqrt{s_{_{NN}}}= 54.4$~GeV.

As pointed out in the previous section the NLO corrections for the 
hyper-skewness ratio $R_{51}^B$ is a factor three smaller than that
for the hyper-kurtosis ratio $R_{62}^B$. Truncation errors for the former 
series are thus expected to be less severe. Furthermore, this ratio will also
be easier to determine experimentally with smaller statistical errors.
It thus would be an important check on the thermodynamic consistency of
higher order cumulants to compare experimental data on $R_{51}^P$ at
$\sqrt{s_{_{NN}}}\ge 54.4$~GeV with the NLO lattice QCD calculations
presented here.

\section{Summary and Conclusions}

We have presented new results on the $\mu_B$-dependence of up to $6^{th}$ 
order cumulants using our latest results on up to $8^{th}$ order cumulants  
calculated at vanishing chemical potentials. 
Using simulation results obtained on lattices of size $32^3\times 8$ and 
$48^3\times 12$ we further presented continuum limit estimates for up to 
$4^{th}$ order 
cumulant ratios. For this analysis we used results from NNLO expansions of 
cumulants in the baryon chemical potential
for strangeness neutral systems, $n_S=0$ at a electric-charge to baryon 
number ratio $n_Q/n_B=0.4$. Systematic effects arising from the truncation
of Taylor series for the skewness and kurtosis
ratios are shown to be small for $\mu_B/T\le 1$, {\it i.e.} for the range
of chemical potentials that can be probed in
heavy ion collisions in a range of beam energies $\sqrt{s_{_{NN}}}\ge 27$~GeV.
A comparison of the results on ratios of up to $4^{th}$ order cumulants of 
net baryon-number fluctuations calculated in equilibrium QCD thermodynamics
with corresponding cumulants of net proton-number fluctuations yields
quite good agreement. This suggests that
the latter are consistent with reflecting the imprint of thermodynamical
fluctuations at a temperature close to but below the pseudo-critical 
temperatures $T_{pc}(\mu_B)$. The particularly good agreement between
lattice QCD calculations and the high statistics experimental data for up 
to fourth order cumulants at $\sqrt{s_{_{NN}}}\ge 54.4$~GeV suggests that 
this conclusion could be further strengthened, would data with
similarly high statistics become available also at other beam 
energies in the range $\sqrt{s_{_{NN}}}\ge 27$~GeV.

We also presented first results from a NLO calculation of $5^{th}$ and $6^{th}$
order cumulants and showed that the hyper-skewness and hyper-kurtosis
ratios $R_{51}^B$ and $R_{62}^B$
are negative at low values of $\mu_B/T$ and temperatures in the vicinity
of $T_{pc}(\mu_B)$. This is at odd with preliminary data obtained 
by the STAR Collaboration at $\sqrt{s_{_{NN}}}\ge 54.4$~GeV for
the $6^{th}$ order cumulant ratio, $R_{62}^P$, of net proton-number 
fluctuations, which is found to be positive and close to unity.
However, on the one hand corrections to the LO result for these cumulants,
calculated in lattice QCD,
are large already at $\mu_B\simeq 0.5$. This makes a calculation of NNLO 
corrections for these cumulants desirable. On the other hand,
the experimental determination of $6^{th}$ order cumulant ratios is known
to require high statistics and current experimental data may be 
statistics limited. We pointed out that a measurement of ratios of $5^{th}$
and $1^{st}$ order cumulants would be very helpful as this ratio can be better 
controlled experimentally and suffer less from truncation effects in NLO 
lattice QCD calculations.

All data from our calculations, presented in the figures of this paper,
can be found at \cite{pubdata}.\href{https://pub.uni-bielefeld.de/record/2941824}{https://pub.uni-bielefeld.de/record/2941824}

\vspace*{0.2cm} 
\noindent
{\it Acknowledgements:} 
This work was supported by: (i) The U.S. Department of Energy, Office of 
Science, Office of Nuclear Physics through the Contract No. DE-SC0012704; 
(ii) The U.S. Department of Energy, Office of Science, Office of Nuclear 
Physics and Office of Advanced Scientific Computing Research within the 
framework of Scientific Discovery through Advance Computing (SciDAC) award 
Computing the Properties of Matter with Leadership Computing Resources; 
(iii) The U.S. Department of Energy, Office of Science, Office of Nuclear 
Physics within the framework of the Beam Energy Scan Theory (BEST) Topical 
Collaboration;
(iv) The U.S. National Science Foundation under award PHY-1812332;
(v) The Deutsche Forschungsgemeinschaft (DFG, German Research Foundation) - Project 
  number 315477589-TRR 211; 
(vi) The grant 05P2018 (ErUM-FSP T01) of the German Bundesministerium f\"ur 
Bildung und Forschung; 
(vii) The European Union H2020-MSCA-ITN-2018-813942(EuroPLEx); 
(viii) The National Natural Science Foundation of China under grant numbers 
11775096 and 11535012 (HTD); 
(ix) The Early Career Research Award of the Science and Engineering Research 
Board of the Government of India (PH); 
(x)  Ramanujan  Fellowship  of  the  Department  of  Science  and Technology,  
Government  of  India (SS).

This research used awards of computer time provided by: 
(i) The INCITE program at Oak Ridge Leadership Computing Facility, a DOE 
Office of Science User Facility operated under Contract No. DE-AC05-00OR22725; 
(ii) The ALCC program at National Energy Research Scientific Computing Center, 
a U.S. Department of Energy Office of Science User Facility operated under 
Contract No. DE-AC02-05CH11231; 
(iii) The INCITE program at Argonne Leadership Computing Facility, a U.S. 
Department of Energy Office of Science User Facility operated under Contract 
No. DE-AC02-06CH11357; 
(iv) The USQCD resources at the Thomas Jefferson National Accelerator Facility.

This research also used computing resources made available through:  
(i) a  PRACE grant at CINECA, Italy; 
(ii) the Gauss Center at NIC-J\"ulich, Germany; 
(iii) the Nuclear Science Computing Center at Central China Normal University 
and 
(iv) the GPU-cluster at Bielefeld University, Germany.

  \newpage

\appendix
\section{Taylor expansion coefficients net baryon-number cumulants}

We give here explicit expressions for the first four expansion coefficients in 
Taylor series for net baryon-number cumulants in strangeness neutral
systems ($n_S=0$) with a fixed ratio of electric charge to baryon number
densities ($n_Q/n_B=0.4$) as defined in Eq.~\ref{chin}. These constraints
determine the strangeness and electric charge chemical potentials 
($\mu_S,\ \mu_Q$) in terms of the baryon chemical potential $\mu_B$
\cite{Bazavov:2017dus},
\begin{eqnarray}
	\hat{\mu}_Q(T,\mu_B) &=& q_1(T)\hat{\mu}_B + q_3(T) \hat{\mu}_B^3+q_5(T) \hat{\mu}_B^5+ \dots \; , \nonumber \\
	\hat{\mu}_S(T,\mu_B) &=& s_1(T)\hat{\mu}_B + s_3(T) \hat{\mu}_B^3+s_5(T) \hat{\mu}_B^5 +..  
\label{qs}
\end{eqnarray}
Explicit expressions for the expansion coefficients $q_i$ and $s_i$ up to 
$i=5$ are given in appendix B of \cite{Bazavov:2017dus}. Results for 
$i=7$ can easily be generated following the procedure outlined in that appendix.

The expansion coefficients of the cumulant series $\chi_n^B(T,\mu_B)$
defined in Eq.~\ref{chin} are given in terms of the expansion coefficients
of the pressure series,
\begin{equation}
	\frac{P}{T^4} = 
	\sum_{i,j,k} \frac{1}{i!j!k!} \chi^{BQS}_{ijk} \hat{\mu}_B^i
	\hat{\mu}_Q^j \hat{\mu}_S^k \; .
	\label{pressurea}
\end{equation}
For $n$ even, one obtains for the expansion coefficients $\tilde{\chi}_n^{B,k}$,
appearing in Eq.~\ref{chin}
\begin{align*}
\tilde{\chi}^{B,0}_{n}&=\chi^{BQS}_{n00}\\
\tilde{\chi}^{B,2}_{n}&=(\chi^{BQS}_{n+2,00}+s_{1}^2 \chi^{BQS}_{n02}+q_{1}^2 \chi^{BQS}_{n20}+2 s_{1} \chi^{BQS}_{n+1,01}\\
&+2 q_{1} \chi^{BQS}_{n+1,10}+2 q_{1} s_{1} \chi^{BQS}_{n11})/2\\
\tilde{\chi}^{B,4}_{n}&=(24 s_{1} s_{3} \chi^{BQS}_{n02}+s_{1}^4 \chi^{BQS}_{n04}
+24 q_{3} s_{1} \chi^{BQS}_{n11}\\
&+24 q_{1} s_{3} \chi^{BQS}_{n11}+4 q_{1} s_{1}^3 \chi^{BQS}_{n13}
+24 q_{1} q_{3} \chi^{BQS}_{n20}\\
&+6 q_{1}^2 s_{1}^2 \chi^{BQS}_{n22}+4 q_{1}^3 s_{1} \chi^{BQS}_{n31}+q_{1}^4 \chi^{BQS}_{n40}\\
&+24 s_{3} \chi^{BQS}_{n+1,01}+4 s_{1}^3 \chi^{BQS}_{n+1,03}
+24 q_{3} \chi^{BQS}_{n+1,10}\\
&+12 q_{1} s_{1}^2 \chi^{BQS}_{n+1,12}+12 q_{1}^2 s_{1} \chi^{BQS}_{n+1,21}+4 q_{1}^3 \chi^{BQS}_{n+1,30}\\
&+6 s_{1}^2 \chi^{BQS}_{n+2,02}+12 q_{1} s_{1} \chi^{BQS}_{n+2,11}
+6 q_{1}^2 \chi^{BQS}_{n+2,20}\\
&+4 s_{1} \chi^{BQS}_{n+3,01}+4 q_{1} \chi^{BQS}_{n+3,10}+\chi^{BQS}_{n+4,00})/24
\end{align*}
\begin{align*}
\tilde{\chi}^{B,6}_{n}&=(360 s_{3}^2 \chi^{BQS}_{n02}+720 s_{1} s_{5} \chi^{BQS}_{n02}\\
&+120 s_{1}^3 s_{3} \chi^{BQS}_{n04}+s_{1}^6 \chi^{BQS}_{n06}+720 q_{5} s_{1} \chi^{BQS}_{n11}\\
&+720 q_{3} s_{3} \chi^{BQS}_{n11}+720 q_{1} s_{5} \chi^{BQS}_{n11}+120 q_{3} s_{1}^3 \chi^{BQS}_{n13}\\
&+360 q_{1} s_{1}^2 s_{3} \chi^{BQS}_{n13}+6 q_{1} s_{1}^5 \chi^{BQS}_{n15}
+360 q_{3}^2 \chi^{BQS}_{n20}\\
&+720 q_{1} q_{5} \chi^{BQS}_{n20}+360 q_{1} q_{3} s_{1}^2 \chi^{BQS}_{n22}+360 q_{1}^2 s_{1} s_{3} \chi^{BQS}_{n22}\\
&+15 q_{1}^2 s_{1}^4 \chi^{BQS}_{n24}
+360 q_{1}^2 q_{3} s_{1} \chi^{BQS}_{n31}+120 q_{1}^3 s_{3} \chi^{BQS}_{n31}\\
&+20 q_{1}^3 s_{1}^3 \chi^{BQS}_{n33}+120 q_{1}^3 q_{3} \chi^{BQS}_{n40}+15 q_{1}^4 s_{1}^2 \chi^{BQS}_{n42}\\
&+6 q_{1}^5 s_{1} \chi^{BQS}_{n51}+q_{1}^6 \chi^{BQS}_{n60}+720 s_{5} \chi^{BQS}_{n+1,01}\\
&+360 s_{1}^2 s_{3} \chi^{BQS}_{n+1,03}+6 s_{1}^5 \chi^{BQS}_{n+1,05}+720 q_{5} \chi^{BQS}_{n+1,10}\\
&+360 q_{3} s_{1}^2 \chi^{BQS}_{n+1,12}+720 q_{1} s_{1} s_{3} \chi^{BQS}_{n+1,12}\\
&+30 q_{1} s_{1}^4 \chi^{BQS}_{n+1,14}+720 q_{1} q_{3} s_{1} \chi^{BQS}_{n+1,21}\\
&+360 q_{1}^2 s_{3} \chi^{BQS}_{n+1,21} +60 q_{1}^2 s_{1}^3 \chi^{BQS}_{n+1,23}+360 q_{1}^2 q_{3} \chi^{BQS}_{n+1,30}\\
&+60 q_{1}^3 s_{1}^2 \chi^{BQS}_{n+1,32}+30 q_{1}^4 s_{1} \chi^{BQS}_{n+1,41}+6 q_{1}^5 \chi^{BQS}_{n+1,50}\\
&+360 s_{1} s_{3} \chi^{BQS}_{n+2,02}
+15 s_{1}^4 \chi^{BQS}_{n+2,04}+360 q_{3} s_{1} \chi^{BQS}_{n+2,11}\\
&+360 q_{1} s_{3} \chi^{BQS}_{n+2,11}+60 q_{1} s_{1}^3 \chi^{BQS}_{n+2,13}
+360 q_{1} q_{3} \chi^{BQS}_{n+2,20}\\
&+90 q_{1}^2 s_{1}^2 \chi^{BQS}_{n+2,22}
+60 q_{1}^3 s_{1} \chi^{BQS}_{n+2,31}+15 q_{1}^4 \chi^{BQS}_{n+2,40}\\
&+120 s_{3} \chi^{BQS}_{n+3,01}+20 s_{1}^3 \chi^{BQS}_{n+3,03}
+120 q_{3} \chi^{BQS}_{n+3,10}\\
&+60 q_{1} s_{1}^2 \chi^{BQS}_{n+3,12}+60 q_{1}^2 s_{1} \chi^{BQS}_{n+3,21}+20 q_{1}^3 \chi^{BQS}_{n+3,30}\\
&+15 s_{1}^2 \chi^{BQS}_{n+4,02}+30 q_{1} s_{1} \chi^{BQS}_{n+4,11}+15 q_{1}^2 \chi^{BQS}_{n+4,20}\\
&+6 s_{1} \chi^{BQS}_{n+5,01}+6 q_{1} \chi^{BQS}_{n+5,10}+\chi^{BQS}_{n+6,00})/720
\end{align*}

For the expansion coefficients of cumulants $\chi_n^B(T,\mu_B)$, with
$n$ odd, one obtains 
\begin{align*}
\tilde{\chi}^{B,1}_{n}&=s_{1} \chi^{BQS}_{n01} + q_{1} \chi^{BQS}_{n10} + \chi^{BQS}_{n+1,00}\\
\tilde{\chi}^{B,3}_{n}&=(6 s_{3} \chi^{BQS}_{n01} + s_{1}^3 \chi^{BQS}_{n03} + 6 q_{3} \chi^{BQS}_{n10}  +  3 q_{1} s_{1}^2 \chi^{BQS}_{n12} \\
&+ 3 q_{1}^2 s_{1} \chi^{BQS}_{n21} + q_{1}^3 \chi^{BQS}_{n30}
+ 3 s_{1}^2 \chi^{BQS}_{n+1,02} + 6 q_{1} s_{1} \chi^{BQS}_{n+1,11} \\
&+ 3 q_{1}^2 \chi^{BQS}_{n+1,20}  + 3 s_{1} \chi^{BQS}_{n+2,01} + 3 q_{1} \chi^{BQS}_{n+2,10} + \chi^{BQS}_{n+3,00})/6\\
\tilde{\chi}^{B,5}_{n}&=(120 s_{5} \chi^{BQS}_{n01} + 60 s_{1}^2 s_{3} \chi^{BQS}_{n03} + s_{1}^5 \chi^{BQS}_{n05} +   120 q_{5} \chi^{BQS}_{n10} \\
&+ 60 q_{3} s_{1}^2 \chi^{BQS}_{n12}  +   120 q_{1} s_{1} s_{3} \chi^{BQS}_{n12} + 5 q_{1} s_{1}^4 \chi^{BQS}_{n14}\\ 
& +   120 q_{1} q_{3} s_{1} \chi^{BQS}_{n21} + 60 q_{1}^2 s_{3} \chi^{BQS}_{n21}  +   10 q_{1}^2 s_{1}^3 \chi^{BQS}_{n23} \\
& + 60 q_{1}^2 q_{3} \chi^{BQS}_{n30}
 +   10 q_{1}^3 s_{1}^2 \chi^{BQS}_{n32} + 5 q_{1}^4 s_{1} \chi^{BQS}_{n41}  \\
&+   q_{1}^5 \chi^{BQS}_{n50} + 120 s_{1} s_{3} \chi^{BQS}_{n+1,02} + 5 s_{1}^4 \chi^{BQS}_{n+1,04}\\ 
& +   120 q_{3} s_{1} \chi^{BQS}_{n+1,11} + 120 q_{1} s_{3} \chi^{BQS}_{n+1,11}  +   20 q_{1} s_{1}^3 \chi^{BQS}_{n+1,13} \\
&+ 120 q_{1} q_{3} \chi^{BQS}_{n+1,20} +   30 q_{1}^2 s_{1}^2 \chi^{BQS}_{n+1,22} + 20 q_{1}^3 s_{1} \chi^{BQS}_{n+1,31} \\ 
&+   5 q_{1}^4 \chi^{BQS}_{n+1,40} + 60 s_{3} \chi^{BQS}_{n+2,01} + 10 s_{1}^3 \chi^{BQS}_{n+2,03} \\
&+   60 q_{3} \chi^{BQS}_{n+2,10} + 30 q_{1} s_{1}^2 \chi^{BQS}_{n+2,12}  
+   30 q_{1}^2 s_{1} \chi^{BQS}_{n+2,21} \\
&+ 10 q_{1}^3 \chi^{BQS}_{n+2,30} + 10 s_{1}^2 \chi^{BQS}_{n+3,02}
 +   20 q_{1} s_{1} \chi^{BQS}_{n+3,11} \\
&+ 10 q_{1}^2 \chi^{BQS}_{n+3,20} + 5 s_{1} \chi^{BQS}_{n+4,01}  +   5 q_{1} \chi^{BQS}_{n+4,10} \\
&+ \chi^{BQS}_{n+5,00})/120\\
\end{align*}
\begin{align*}
\tilde{\chi}^{B,7}_{n}&=(5040 s_{7} \chi^{BQS}_{n01} + 2520 s_{1} s_{3}^2 \chi^{BQS}_{n03}
+   2520 s_{1}^2 s_{5} \chi^{BQS}_{n03} \\
&+ 210 s_{1}^4 s_{3} \chi^{BQS}_{n05}  +   s_{1}^7 \chi^{BQS}_{n07} + 5040 q_{7} \chi^{BQS}_{n10} \\
&+ 2520 q_{5} s_{1}^2 \chi^{BQS}_{n12}
+   5040 q_{3} s_{1} s_{3} \chi^{BQS}_{n12} + 2520 q_{1} s_{3}^2 \chi^{BQS}_{n12}  \\
&+   5040 q_{1} s_{1} s_{5} \chi^{BQS}_{n12} + 210 q_{3} s_{1}^4 \chi^{BQS}_{n14}
+   840 q_{1} s_{1}^3 s_{3} \chi^{BQS}_{n14} \\
&+ 7 q_{1} s_{1}^6 \chi^{BQS}_{n16}  +   2520 q_{3}^2 s_{1} \chi^{BQS}_{n21} + 5040 q_{1} q_{5} s_{1} \chi^{BQS}_{n21}\\ 
& +   5040 q_{1} q_{3} s_{3} \chi^{BQS}_{n21} + 2520 q_{1}^2 s_{5} \chi^{BQS}_{n21}  +   840 q_{1} q_{3} s_{1}^3 \chi^{BQS}_{n23} \\
&+ 1260 q_{1}^2 s_{1}^2 s_{3} \chi^{BQS}_{n23}
+   21 q_{1}^2 s_{1}^5 \chi^{BQS}_{n25} + 2520 q_{1} q_{3}^2 \chi^{BQS}_{n30} \\
&+   2520 q_{1}^2 q_{5} \chi^{BQS}_{n30} + 1260 q_{1}^2 q_{3} s_{1}^2 \chi^{BQS}_{n32}\\ 
& +   840 q_{1}^3 s_{1} s_{3} \chi^{BQS}_{n32} + 35 q_{1}^3 s_{1}^4 \chi^{BQS}_{n34}  +   840 q_{1}^3 q_{3} s_{1} \chi^{BQS}_{n41}\\ 
&+ 210 q_{1}^4 s_{3} \chi^{BQS}_{n41}
+   35 q_{1}^4 s_{1}^3 \chi^{BQS}_{n43} + 210 q_{1}^4 q_{3} \chi^{BQS}_{n50}\\
&+   21 q_{1}^5 s_{1}^2 \chi^{BQS}_{n52} + 7 q_{1}^6 s_{1} \chi^{BQS}_{n61}
+   q_{1}^7 \chi^{BQS}_{n70}
+ 2520 s_{3}^2 \chi^{BQS}_{n+1,02} \\
&+ 5040 s_{1} s_{5} \chi^{BQS}_{n+1,02}  +   840 s_{1}^3 s_{3} \chi^{BQS}_{n+1,04}
+ 7 s_{1}^6 \chi^{BQS}_{n+1,06} \\
&+   5040 q_{5} s_{1} \chi^{BQS}_{n+1,11} + 5040 q_{3} s_{3} \chi^{BQS}_{n+1,11} 
+   5040 q_{1} s_{5} \chi^{BQS}_{n+1,11} \\
&+ 840 q_{3} s_{1}^3 \chi^{BQS}_{n+1,13} +   2520 q_{1} s_{1}^2 s_{3} \chi^{BQS}_{n+1,13}
+ 42 q_{1} s_{1}^5 \chi^{BQS}_{n+1,15} \\
&+   2520 q_{3}^2 \chi^{BQS}_{n+1,20} + 5040 q_{1} q_{5} \chi^{BQS}_{n+1,20}
+   2520 q_{1} q_{3} s_{1}^2 \chi^{BQS}_{n+1,22}\\
&+ 2520 q_{1}^2 s_{1} s_{3} \chi^{BQS}_{n+1,22}  +   105 q_{1}^2 s_{1}^4 \chi^{BQS}_{n+1,24}\\ 
&+ 2520 q_{1}^2 q_{3} s_{1} \chi^{BQS}_{n+1,31}
+   840 q_{1}^3 s_{3} \chi^{BQS}_{n+1,31} + 140 q_{1}^3 s_{1}^3 \chi^{BQS}_{n+1,33} \\
&+   840 q_{1}^3 q_{3} \chi^{BQS}_{n+1,40} + 105 q_{1}^4 s_{1}^2 \chi^{BQS}_{n+1,42} + 42 q_{1}^5 s_{1} \chi^{BQS}_{n+1,51}\\
& +   7 q_{1}^6 \chi^{BQS}_{n+1,60} + 2520 s_{5} \chi^{BQS}_{n+2,01}  +   1260 s_{1}^2 s_{3} \chi^{BQS}_{n+2,03} \\
&+ 21 s_{1}^5 \chi^{BQS}_{n+2,05} + 2520 q_{5} \chi^{BQS}_{n+2,10}
+   1260 q_{3} s_{1}^2 \chi^{BQS}_{n+2,12} \\
&+ 2520 q_{1} s_{1} s_{3} \chi^{BQS}_{n+2,12}
+   105 q_{1} s_{1}^4 \chi^{BQS}_{n+2,14} + 2520 q_{1} q_{3} s_{1} \chi^{BQS}_{n+2,21}\\ 
& +   1260 q_{1}^2 s_{3} \chi^{BQS}_{n+2,21} + 210 q_{1}^2 s_{1}^3 \chi^{BQS}_{n+2,23}  +   1260 q_{1}^2 q_{3} \chi^{BQS}_{n+2,30} \\
&+ 210 q_{1}^3 s_{1}^2 \chi^{BQS}_{n+2,32}
+   105 q_{1}^4 s_{1} \chi^{BQS}_{n+2,41} + 21 q_{1}^5 \chi^{BQS}_{n+2,50}\\
&+   840 s_{1} s_{3} \chi^{BQS}_{n+3,02} + 35 s_{1}^4 \chi^{BQS}_{n+3,04}
+   840 q_{3} s_{1} \chi^{BQS}_{n+3,11} \\
&+ 840 q_{1} s_{3} \chi^{BQS}_{n+3,11}  +   140 q_{1} s_{1}^3 \chi^{BQS}_{n+3,13} + 840 q_{1} q_{3} \chi^{BQS}_{n+3,20}\\ 
& +   210 q_{1}^2 s_{1}^2 \chi^{BQS}_{n+3,22} + 140 q_{1}^3 s_{1} \chi^{BQS}_{n+3,31}  +   35 q_{1}^4 \chi^{BQS}_{n+3,40} \\
&+ 210 s_{3} \chi^{BQS}_{n+4,01} + 35 s_{1}^3 \chi^{BQS}_{n+4,03}
+   210 q_{3} \chi^{BQS}_{n+4,10} \\
&+ 105 q_{1} s_{1}^2 \chi^{BQS}_{n+4,12}  +   105 q_{1}^2 s_{1} \chi^{BQS}_{n+4,21} + 35 q_{1}^3 \chi^{BQS}_{n+4,30} \\
&+ 21 s_{1}^2 \chi^{BQS}_{n+5,02}
+   42 q_{1} s_{1} \chi^{BQS}_{n+5,11} + 21 q_{1}^2 \chi^{BQS}_{n+5,20} \\
&+ 7 s_{1} \chi^{BQS}_{n+6,01}  +   7 q_{1} \chi^{BQS}_{n+6,10} + \chi^{BQS}_{n+7,00})/5040
\end{align*}

\end{document}